\documentclass[11pt]{article}

\usepackage{authblk}
\usepackage{graphicx}
\usepackage{pdfsync}
\usepackage{hyperref}
\usepackage{statex2}
 \usepackage{forest}
 \usepackage{tikz}
\usepackage{caption}
\usepackage{subcaption}

\newcommand*{\bart}[1]{\mathrm{BART}\wrap[()]{#1}}
\newcommand*{\citep}[1]{(\cite{#1})} % these two commands are a workaround
\newcommand*{\citet}[1]{\cite{#1}} % What is the right way to do this?
\renewcommand*{\Pr}[2][]{\mathrm{Pr}\ifthenelse{\equal{#1}{}}{}{_{#1}}\wrap[()]{#2}}

\begin{document}
\title{Nonparametric competing risks analysis using\\ Bayesian Additive 
Regression Trees (BART)}
 \author[1]{Rodney Sparapani}
 \author[1]{Brent R. Logan}
 \author[2]{Robert E. McCulloch}
 \author[1]{Purushottam W. Laud}
 \affil[1]{Division of Biostatistics, Medical College of Wisconsin}
 \affil[2]{School of Mathematical and Statistical Sciences, Arizona State University}
\date{}                     %% if you don't need date to appear
\setcounter{Maxaffil}{0}
\renewcommand\Affilfont{\itshape\small}

\maketitle

\begin{abstract}
Many time-to-event studies are complicated by the presence of competing risks.  Such data are often analyzed using Cox models for the cause specific hazard function or Fine-Gray models for the subdistribution hazard.  In practice regression relationships in competing risks data with either strategy are
often complex and may include nonlinear functions of covariates,
interactions, high-dimensional parameter spaces and nonproportional cause specific or
subdistribution hazards. Model misspecification can lead to poor
predictive performance.  To address these issues, we propose a novel approach to flexible prediction modeling of competing risks data using Bayesian Additive Regression Trees (BART).  We study the simulation performance in two-sample scenarios as well as a complex regression setting, and benchmark its performance against standard regression techniques as well as random survival forests.  We illustrate the use of the proposed method on a recently published study of patients undergoing hematopoietic stem cell transplantation.    
\end{abstract} 

\maketitle
  
\section{Introduction}

Many time-to-event studies in biomedical applications are complicated
by the presence of competing risks: a patient can fail from one of
several different causes, and the occurrence of one kind of failure
precludes the observation of another kind.  With little loss in
generality, the event kinds are often categorized as a cause of interest
(cause 1) or a competing event from any other cause (cause 2).  If a
patient experiences the cause 2 competing event, they are no longer at
risk of experiencing the cause 1 event after the competing event time.
This is different from censoring, where a patient who is censored or lost to
follow up is still potentially able to experience either event
kind after the censoring time.  Several approaches to modeling such
data have been proposed which target different parameters.
Historically, Cox regression models were used to model each
cause-specific hazard function $\lambda_k(t)$ as a specified function
of covariates (\cite{PrenKalb78}).  However, unlike with survival
analysis, there is not a one-to-one correspondence between the cause
specific hazard
function for cause 1 and the cumulative incidence function $F_1(t)$
which is defined as the probability of failing from cause 1 before
time $t$.  In fact, $F_1(t)$ depends on the cause specific hazards for
all failure causes.  Indirect inference on the cumulative incidence
function can be obtained by combining the estimates of the cause
specific hazard functions as in \cite{AndeBorg93} (pp. 512--515).
Alternatively, Fine and Gray\cite{FineGray99} proposed a proportional
subdistribution hazards regression model leading to direct
inference on the cumulative incidence function.  Others 
have proposed regression
methods that more directly model the cumulative incidence through a link
function \cite{KleiAnde05,ScheZhan08}.

In practice, regression relationships in competing risks data are
often complex. These can involve nonlinear functions of covariates,
interactions, high-dimensional parameter spaces and nonproportional cause-specific or
subdistribution hazards. Model misspecification can lead to poor
predictive performance.  Several solutions have been proposed to
address these complexities and focus on improved prediction in the
survival setting.  In the survival data setting without competing risks, 
these include variable selection using lasso-type penalization 
\citet{Tibs97,ParkHast07,ZhanLu07}, flexible prediction models using boosting
with Cox-gradient descent \citep{LiLuan06,MaHuan06}, random survival
forests \citep{IshwKoga08} and our previous work with Bayesian
Additive Regression Trees (BART) described further below
\cite{SparLoga16}. Support vector machines
\citep{VanBPelc10} have also been used in the survival setting to determine
a function of covariates which is concordant with the observed failure
times; however, this only leads to a ranking of risk profiles and does
not directly provide predictions of survival probabilities that are
often of clinical interest. 

In the competing risks setting, there are fewer modeling approaches proposed 
to alleviate the above mentioned modeling concerns.  Penalized
variable selection for the Fine and Gray model \citep{FuPari17,AhnBane17} 
and an extension of random survival forests 
\cite{IshwGerd14} have been considered.  In this article, we describe
a new approach to flexible prediction modeling of competing risks data using BART
that allows for complex functional forms of the covariates, does not
require restrictive proportional or subdistribution hazards
assumptions, can account for high-dimensional parameter spaces, and can
accomplish inference on a wide variety of model functionals of interest at
little additional overhead in mathematical or computational effort.

BART \cite{ChipGeor10} is an ensemble of trees model which has been shown
to be efficient and flexible with performance comparable to or better
than competitors such as boosting, lasso, MARS, neural nets and random
forests.  In addition, recent modifications to the BART prior have
been proposed that maintain excellent out-of-sample predictive
performance even when a large number of additional irrelevant
regressors are added \cite{Line17}.  Finally, the
Bayesian framework naturally leads to quantification of
uncertainty for statistical inference of the cumulative incidence
functions or other related quantities.  Because of its tree-based
structure, BART can effectively address interactions among variables
including, in our case, interactions with time to allow for
nonproportional hazards.  

Our method re-expresses the nonparametric likelihood for competing
risks data in a form suitable for BART. We examine two different ways
of re-expressing this likelihood that leads to two different BART
competing risks models.  In both cases, two BART models are needed
to adequately reflect the relationships between covariates and the
relevant model parameters.  However, we can employ existing
BART software by suitably partitioning the data for each
BART component.

We present our work in the following sequence. In Section 2, we
review BART methodology, along with our previous extension of
BART to survival data.  In Section 3 we propose two ways of adapting BART to
competing risks analysis. Section 4 studies the performance of the
proposed methods including examining various proportional and
subdistribution hazards models in a two sample setting. We also
demonstrate the model's ability to accommodate data from complex
regression models.  In Section 5, we present a health care application that
illustrates the advantages of the proposed methodology.  We summarize our contribution and describe some planned future developments in Section 6.

\section{Background in BART methodology}\label{Methodology}

As BART is based on a collection of regression tree models, we begin
with a simple example of a regression tree model. We then describe how
BART uses an ensemble of regression tree models for a numeric outcome.
We discuss how the BART model for a numeric outcome is
augmented to model a binary outcome.  This binary BART model will be
directly utilized in our competing risk models by the transformation of
the survival data into a sequence of binary indicators.  Finally, we review how the BART model can be adapted to handle high dimensional predictors.  

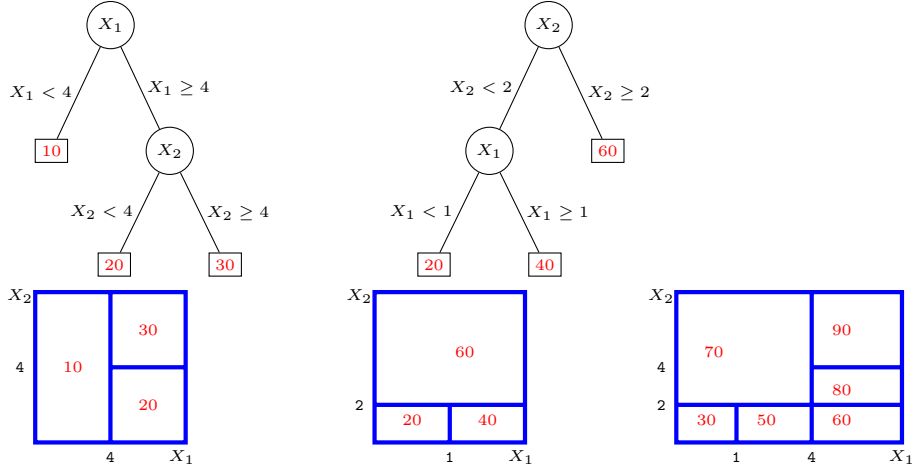
\begin{figure}

%\centering
\medskip
{ \tiny
%  \hspace{-4cm}
  \begin{forest}
  for tree={font=\tiny,l sep =4em, s sep=4em, anchor=center} 
  [$X_1$, circle, draw,
    [\textcolor{red}{$10$},rectangle,draw, edge label={node[midway,left]{$X_1<4$}}]
    [$X_2$, edge label={node[midway,right]{$X_1\ge 4$}}, circle, draw,  
      [\textcolor{red}{$20$}, rectangle,draw,edge label={node[midway,left]{$X_2<4$}}]
      [ \textcolor{red}{$30$}, rectangle,draw,edge label={node[midway,right]{$X_2\ge 4$}}]]]
  \end{forest}
}
{ \tiny
  \hspace{1cm}
  \begin{forest}
  for tree={font=\tiny,l sep =4em, s sep=4em, anchor=center} 
  [$X_2$, circle, draw,
    [$X_1$, edge label={node[midway,left]{$X_2< 2$}}, circle, draw,  
     [\textcolor{red}{$20$}, rectangle,draw,edge label={node[midway,left]{$X_1<1$}}]
     [ \textcolor{red}{$40$}, rectangle,draw,edge label={node[midway,right]{$X_1\ge 1$}}]]
    [\textcolor{red}{$60$},rectangle,draw, edge label={node[midway,right]{$X_2\ge 2$}}]]
  \end{forest}
}

\hspace{-0.2cm}
\begin{tikzpicture}[num/.style={font=\tiny\bfseries\ttfamily}]
\draw[ultra thick,blue] (0,0) rectangle (2,2)
(1,0) -- (1,2)
(1,1) -- (2,1); 
\node[num] at (1,-0.2){4};
\node[num] at (-0.2,1){4} ;
\node[num] at (1.95,-0.2){$X_1$};
\node[num] at (-0.2,1.95){$X_2$} ;  
\node[num] at (0.5,1){\textcolor{red}{$10$}} ;
\node[num] at (1.5,0.5){\textcolor{red}{$20$}} ;
\node[num] at (1.5,1.5){\textcolor{red}{$30$}} ;
\end{tikzpicture}
\hspace{1.5cm}
\begin{tikzpicture}[num/.style={font=\tiny\bfseries\ttfamily}]
\draw[ultra thick,blue] (0,0) rectangle (2,2)
(0,0.5) -- (2,0.5)
(1,0) -- (1,0.5); 
\node[num] at (1,-0.2){1};
\node[num] at (-0.2,0.5){2} ;
\node[num] at (1.95,-0.2){$X_1$};
\node[num] at (-0.2,1.95){$X_2$} ;  
\node[num] at (0.5,0.3){\textcolor{red}{$20$}} ;
\node[num] at (1.5,0.3){\textcolor{red}{$40$}} ;
\node[num] at (1.2,1.2){\textcolor{red}{$60$}} ;
\end{tikzpicture} 
\hspace{1cm}
\begin{tikzpicture}[num/.style={font=\tiny\bfseries\ttfamily}]
\draw[ultra thick,blue] (0,0) rectangle (3,2)
(0,0.5) -- (3,0.5)
(0.8,0) -- (0.8,0.5)
(1.8,0) -- (1.8,2)
(1.8,1) -- (3,1); 
\node[num] at (1.8,-0.2){4};
\node[num] at (-0.2,1){4}; 
\node[num] at (0.8,-0.2){1};
\node[num] at (-0.2,0.5){2} ;
\node[num] at (3,-0.2){$X_1$};
\node[num] at (-0.2,1.95){$X_2$} ;  

\node[num] at (0.4,0.3){\textcolor{red}{$30$}} ;
\node[num] at (0.5,1.2){\textcolor{red}{$70 $}} ;
\node[num] at (2.2,0.7){\textcolor{red}{$80$}} ;
\node[num] at (1.2,0.3){\textcolor{red}{$50$}}; 
\node[num] at (2.2,0.3){\textcolor{red}{$60$}} ;
\node[num] at (2.2,1.5){\textcolor{red}{$90$}} ;
\end{tikzpicture}
\caption{Two trees (left and center), and their sum (right), with a
  two-dimensional covariate.  Each tree is represented both in
  leaf/node form as well as rectangular partition. \label{twotrees}}
\end{figure}

Suppose $y_i$ represents the numeric outcome for individual $i$, and
$\bm{x}_i$ is a vector of covariates with the regression relationship
$y_i=g(\bm{x}_i; T, M)+\epsilon_i$ where $i=1, \dots, N$.
Notationally, $g(\bm{x}_i; T, M)$ is a binary tree function with
components $T$ and $M$ that can be described as follows.  $T$ denotes
the tree structure consisting of two sets of nodes: interior branches
and terminal leaves.  Each branch is a decision rule that is a binary
split based on a single covariate.  $M=\{\mu_1, \dots, \mu_b \}$ is
made up of the function values of the leaves.  Each leaf is a numeric
value: the value being the corresponding output of $g$ when the branch
rules applied to $\bm{x}_i$ uniquely determine the branch ``climbing''
route to a single leaf.  Examples of two trees are shown in Figure
\ref{twotrees} wherein branches appear as circles, and leaves as
rectangles.  Trees effectively partition the covariate space into
rectangular regions, and these alternative representations are also
shown in the figure.

BART employs an ensemble of such trees in an additive fashion, i.e.,
it is the sum of $m$ trees where $m$ is typically large such as 50,
100 or 200.  Figure \ref{twotrees} shows a simple example of
adding two trees.  Note this sum of trees leads to a finer rectangular
partition of the covariate space compared to each individual tree;
here the value in each rectangular region is the sum of the terminal
nodes in each tree corresponding to that region.  The model can be
represented as:
\begin{equation}\label{eq:bartmodel}
\left. \begin{array}{rcl}
y_i &=&\mu_0+ f(\bm{x}_i)+\epsilon_i \where \epsilon_i \iid \N{0}{\sd^2} \\
f(\bm{x}_i)&=&\sum_{j=1}^m g(\bm{x}_i; T_j, M_j) 
\end{array}\right\}
\end{equation}
where $\mu_0$ is typically set to $\bar{y}$.  
To proceed with the Bayesian paradigm, we need priors for the unknown parameters.  We specify the prior for the error variance as $\sd^2 \prior \nu \lambda \IC{\nu}$; details on specification of the hyperparameters $\nu$ and $\lambda$ are discussed in \cite{ChipGeor10}.  And, notationally, 
we specify the prior for the unknown function, $f$, as:
\begin{equation}\label{eq:bartprior}
f   \prior \bart{m, \mu_0, \tau, \alpha, \gamma} \ 
\end{equation}
and describe it as made up of two components: a prior on the
complexity of each tree, $T_j$, and a prior on its terminal nodes,
$M_j|T_j$.  Using the Smith-Gelfand generic bracket notation
\citep{GelfSmit90} as a shorthand for writing a probability density or
conditional density, we write
$\wrap{f} = \prod_j \wrap{T_j} \wrap{M_j|T_j} $\ . Following
\citep{ChipGeor10}, we partition $\wrap{T_j}$ into 3 components: the
tree structure, or process by which we build a tree creating branches;
the choice of a splitting covariate given a branch; and the choice of
cutpoint given the covariate for that branch.  The probability of a
node being a branch vs.\ a leaf is defined by describing the
probabilistic process by which a tree is grown.  We start with a tree
that is just a single node, or root, and then randomly ``grow'' it
into a branch (with two leaves) by the probability
$\alpha (1+d)^{-\gamma}$ where $d$ represents the branch depth,
$\alpha \in (0, 1)$ and $\gamma \ge 0$.  We assume that the choice of
a splitting covariate given a branch, and the choice of a cutpoint
value given a covariate and a branch, are both uniform.  We then use
the prior $\wrap{M_j| T_j} = \prod_{\ell=1}^{b_j} \wrap{\mu_{j\ell}}$
where $b_j$ is the number of leaves for tree $j$ and
$\mu_{j\ell} \prior\N{\mu_0=0}{\frac{\tau^2}{m}}$.  Here $\tau=\frac{0.5}{\kappa}$ is parametrized in terms of a tuning parameter $\kappa$ with default value of $\kappa=2$ recommended in \cite{ChipGeor10} and used in the {\bf
  BART} R package \citep{McCuSpar18}).  This gives
$f(\bm{x}) \sim N(0, \tau^2)$ for any $\bm{x}$ since $f(\bm{x})$ is
the sum of $m$ independent Normals.  Along with centering of the
outcome, these default prior parameters are specified such that
each tree is a ``weak learner'' playing only a small part in the
ensemble; more details on this can be found in \cite{ChipGeor10}.

For data sets with a large number of covariates, $P$, Linero
\citep{Line17} proposed replacing the uniform prior for selecting a
covariate with a sparse prior.  We refer to this alternative as the
DART prior (the ``D'' is a mnemonic reference to the Dirichlet
distribution).  We represent the probability of variable selection via
a sparse Dirichlet prior as
$\wrap{s_1, \dots, s_P} \prior \Dir{\theta/P, \dots, \theta/P}$ rather
than the uniform probability $1/P$.  The prior parameter $\theta$ can
be fixed or random.  Linero \citep{Line17} recommends that $\theta$ is
random and specified via
$\frac{\theta}{\theta+\rho} \prior \Bet{a}{b}$ with the following
sparse settings: $\rho=P$, $a=0.5$ and $b=1$.  The distribution of
$\theta$, especially the parameters $\rho$ and $a$, control the degree
of sparsity: $a=1$ is not sparse while $a=0.5$ is sparse and
further sparsity can be achieved by setting $\rho<P$.
%$a=0.5$ induces a sparse posture while $a=1$ is not
%sparse and similar to a uniform prior with probability
%$1/P$.  % Linero
% \citep{Line17} shows that this
This Dirichlet sparse prior helps the BART %Bayesian
%tree-based models 
model naturally adapt to sparsity when $P$ is large; both in terms of
improving predictive performance as well as identifying
important predictors in the model. Note that alternative variable
selection methods exist for BART such as
%the average use per splitting rule are 
a permutation-based approach due to Bleich and Kapelner
\citep{BleiKape14} that is available in the {\bf bartMachine} R package
\citep{KapeBlei14}; however, we focus on incorporating the
Dirichlet sparse prior, as an option, into BART in subsequent
competing risks models.
 
To apply the BART model to a binary outcome, we use a probit
transformation
$$
Pr(y=1 \vert \bm{x}) \equiv p(\bm{x})=\Phi(\mu_0+f(\bm{x}))
$$
where $\Phi$ is the standard normal cumulative distribution function
and $f \sim \bart{m, \mu_0, \tau, \alpha, \gamma} $.  To estimate this model, we use the approach
of Albert and Chib \cite{AlbeChib93} and augment the model with latent
variables $z_i$:
\begin{equation}\label{disp:binarybartmodel}
\begin{array}{rcl}
y_i & = & I_{z_i \ge 0} \\
z_i & = & \mu_0 + f(\bm{x}_i) + \epsilon_i \\
f(\bm{x}_i)&=&\sum_{j=1}^m g(\bm{x}_i; T_j, M_j) \\
f & \prior & \bart{m, \mu_0, \tau, \alpha, \gamma}
\end{array}
\end{equation}
where the indicator function $I_{z \ge 0}$ is one if $z \ge 0$, zero
otherwise; and $\epsilon_i \iid N(0,1)$.  The Albert and Chib method
provides draws of $f$ from the posterior via Gibbs sampling, i.e.,
draw $z \vert f$, $f \vert z$, etc.

The model just described can be readily estimated using existing
software for binary BART. It provides inference for the function
$f(\bm{x})$ through Markov Chain Monte Carlo (MCMC) draws of $f$ from
which the corresponding success probabilities,
$p(\bm{x})=\Phi(\mu_0+f(\bm{x}))$, are readily obtained.  Here $\mu_0$
is typically set to $\Phi^{-1}(\bar{y})$.
% is a tuning parameter to be chosen.  For example, setting $\mu_0=0$
% centers the prior for $p(\bm{x})$ at 0.5 since the BART prior for $f$
% is centered at 0; however, the BART model is fairly robust to
% different prior values of $\mu_0$ given sufficient data.
In the
binary probit case, we let $\tau=\frac{3}{\kappa}$, so that there is 0.95 prior probability
that $f(\bm{x})$ is in the interval $(-3,\ 3)$ giving a reasonable
range of values for $p(\bm{x})$.  Note that Logistic latents, rather
than Normal latents,
could also be used for the binary outcome setting, and a Logistic
implementation is also available in the {\bf BART} package.  However,
because we are doing a prediction model and not focusing on parameter
estimates like odds ratios, it is unclear whether probit or Logistic is
more useful, so we have proceeded with the simpler and more
computationally efficient probit framework.

Sparapani et al.\ \citep{SparLoga16} adapted binary probit BART to the
survival setting using discrete-time survival analysis \citep{Fahr98}.
We review this in detail, since a similar discrete-time approach is
used here for the competing risks setting.  Survival data are
typically represented as ($t_i, \delta_i, \bm{x}_i$) where $t_i$ is
the event time, $\delta_i$ is an indicator distinguishing events
($\delta=1$) from right-censoring ($\delta=0$), $\bm{x}_i$ is a vector
of covariates, and $i=1, \dots, N$ indexes subjects.  We denote the
$J$ distinct event and censoring times by
$0<t_{(1)}< \dots< t_{(J)}<\infty$ thus taking $t_{(j)}$ to be the
$j^{th}$ order statistic among distinct observation times and, for
convenience, $t_{(0)}=0$. Now consider event indicators $y_{ij}$ for
each subject $i$ at each distinct time $t_{(j)}$ up to and including
the subject's observation time $t_i=t_{(n_i)}$ with
$n_i=\#\{j:t_{(j)}\leq t_i\}$ or $n_i=\arg \max_j\{t_{(j)}\leq t_i\}$.
This means $y_{ij}=0$ if $j<n_i$ and $y_{in_i}=\delta_i$. We then
denote by $p_{ij}$ the probability of an event at time $t_{(j)}$
conditional on no previous event. The likelihood has the form
\begin{equation}\label{eq:ygivenp}
L(p|\bm{y})\ = \prod_{i=1}^N\prod_{j=1}^{n_i} p_{ij}^{y_{ij}} 
(1-p_{ij})^{1-y_{ij}}\ .
\end{equation}
where the product over $j$ is a result of the definition of $p_{ij}$'s
as conditional probabilities, and not the consequence of an assumption
of independence.  Since this likelihood has the form of a binary
likelihood for $y_{ij}$,
%$i=1,\ldots,n;j=1,\ldots,n_i$, 
we can apply the probit BART model where
$p_{ij}=\Phi(\mu_0+f(t_{(j)},\bm{x}_i))$.  Note the incorporation of
$t$ into the BART function $f(t,x)$ allows the conditional probabilities
to be time-varying, similar to a nonproportional hazards model.

With the data prepared as described above, the BART model for binary
data treats the conditional probability of the event in an interval,
given no events in preceding intervals, as a nonparametric function of
the time $t$ and the covariates $\bm{x}$. Conditioned on the data, the
algorithm in the {\bf BART} package \citep{McCuSpar18} generates
samples, each containing $m$ trees, from the posterior distribution of
$f$. For any $t$ and $\bm{x}$ then, we can obtain posterior samples of
\[ p(t,\bm{x})=\Phi(\mu_0+f(t,\bm{x})) \]
and the survival function 
\[ S(t_{(j)}|\bm{x})=Pr(t>t_{(j)}|\bm{x})=\prod_{l=1}^j
(1-p(t_{(l)},\bm{x})), j=1,\ldots,k\ . \]

BART models with multiple covariates do not directly provide a summary
of the marginal effect for a single covariate, or a subset of
covariates, on the outcome.  Marginal effect summaries are generally a
challenge for nonparametric regression and/or black-box models.  We
use Friedman's partial dependence function \citep{Frie01} with BART to
summarize the marginal effect due to a subset of the covariates,
$\bm{x}_S$, by aggregating over the complement covariates, $\bm{x}_C$,
i.e., $\bm{x} =\wrap{\bm{x}_S,\bm{x}_C}$.  The marginal dependence
function is defined by fixing $\bm{x}_S$ while aggregating over the
observed settings of the complement covariates in the cohort as
follows.
\begin{align}
\label{eq:pdcif}f_S(\bm{x}_S)&={N^{-1}}\sum_{i=1}^N f(\bm{x}_S,\bm{x}_{iC})
\end{align}

Consider the marginal survival function:
$S_S(t|\bm{x}_S) = {N^{-1}} \sum_i S(t|\bm{x}_S,\bm{x}_{iC})$.
Other marginal functions can be obtained in a similar fashion.
Marginal estimates can be derived via functions of the posterior
samples such as means, quantiles, etc.

\section{Competing Risks using BART}

Competing risks data are typically represented as
($t_i, \delta_i, \epsilon_i,\bm{x}_i$) where $\epsilon_i \in \{1,2\}$
denotes the event cause and, similar to before, $t_i$ is the time to
the event or censoring time, $\delta_i$ is an indicator distinguishing
events ($\delta=1$) from right-censoring ($\delta=0$), $\bm{x}_i$ is a
vector of covariates, and $i=1, \dots, N$ indexes subjects.

As before, we denote the $J$ distinct event and censoring times by
$0<t_{(1)}< \dots< t_{(J)}<\infty$, and let $n_i=\arg \max_j\{t_{(j)}\leq t_i\}$. The simplest way of representing
the discrete time competing risks model is through a sequence of
multinomial events $y_{ijk}=I(t_i=t_{(j)},\epsilon_i=k)$, $i=1,\ldots,N;j=1,\ldots,n_i;k=1,2$, and their
corresponding conditional probabilities
$p_{ijk}=P(t=t_{(j)},\epsilon_i=k|t_i\geq t_{(j)})$, which is
interpreted as the probability of an event of cause $k$ at time
$t_{(j)}$ given that the patient is still at risk (has not yet
experienced either cause of event).  Now, by successfully conditioning
over time, we can write the likelihood as
\begin{equation}\label{eq:multinomial}
L(p|\bm{y})\ = \prod_{i=1}^N\prod_{j=1}^{n_i} 
p_{ij1}^{y_{ij1}}p_{ij2}^{y_{ij2}}(1-p_{ij1}-p_{ij2})^{1-y_{ij1}-y_{ij2}}.
\end{equation}
Since this likelihood matches that of a set of independent multinomial
observations, one could directly apply BART models to the multinomial
probabilities \citep{Murr17}.
%{\bf is there a published reference for this?? software??}.  
However, multinomial BART implementations are not as widely available,
and their current approaches require estimation of the same number of
BART functions as multinomial categories.  We propose two alternative
representations of the likelihood that facilitate direct use of the
more prevalent binary probit BART implementations.  Furthermore, our
proposals are more computationally efficient by utilizing fewer BART
functions to model the outcomes (two BART functions instead of three
for a standard competing risk framework with two competing events).

\subsection{Method 1}
In this method, we re-write the likelihood as 

\begin{eqnarray}\label{eq:survplusbinary}
L(p|\bm{y})\ &=& \prod_{i=1}^N\prod_{j=1}^{n_i} \left(\frac{p_{ij1}}{p_{ij1}+p_{ij2}}\right)^{y_{ij1}}  \left(\frac{p_{ij2}}{p_{ij1}+p_{ij2}}\right)^{y_{ij2}} \notag \\
& &\times(p_{ij1}+p_{ij2})^{y_{ij1}+y_{ij2}}(1-p_{ij1}-p_{ij2})^{1-y_{ij1}-y_{ij2}} \notag \\
&=& \prod_{i=1}^N\prod_{j=1}^{n_i} p_{ij\cdot}^{y_{ij\cdot}} (1-p_{ij\cdot})^{1-y_{ij\cdot}} \prod_{i:\delta_i=1} \psi_{i}^{u_{i}}(1-\psi_{i})^{1-u_{i}} 
\end{eqnarray}
where $p_{ij\cdot}=p_{ij1}+p_{ij2}$, $y_{ij\cdot}=y_{ij1}+y_{ij2}$,
$u_i=I(\epsilon_i=1)$ and
$\psi_{i}=P(\epsilon_i=1|t_i,\delta_i=1)$.  This likelihood separates
into two binary likelihoods, so that we can fit two separate BART probit
models for $p_{ij\cdot}$ and $\psi_{i}$, using the corresponding
binary observations $y_{ij\cdot}$ and $u_{i}$ respectively.
Specifically, we assume
\begin{equation}\label{disp:bartmodel1}
\begin{array}{rcl}
y_{ij\cdot} & = & I_{z_{ij} \ge 0} \\
z_{ij} & = & \mu_{y} + f_{y}(t_{(j)},\bm{x}_{i}) + e_{ij} \\
f_{y} & \sim & \bart{m, \mu_{y}, \tau, \alpha, \gamma}
\end{array}
\end{equation}
for the first model and 
\begin{equation}\label{disp:bartmodel1}
\begin{array}{rcl}
u_{i} & = & I_{\tilde{z}_{i} \ge 0} \\
\tilde{z}_{i} & = & \mu_{u} + f_u(t_i,\bm{x}_{i}) + \tilde{e}_{i} \\
f_{u} & \sim & \bart{m, \mu_{u}, \tau, \alpha, \gamma}
\end{array}
\end{equation}
for the second model.  Conceptually, the first BART model is
equivalent to a BART survival model for the time to the first event,
while the latter BART model accounts for the probability of the event
being of failure cause 1 given that an event occurs.

The algorithms in existing BART software provide for samples from the
posterior distribution of $f_y$ and $f_u$ given the data.
Similarly, samples from the posterior distribution of
$p_y(t,\bm{x})=\Phi(\mu_{y}+f_{y}(t,\bm{x}))$ and
$\psi(t,\bm{x})=\Phi(\mu_{u}+f_{u}(t,\bm{x}))$.  Inference on the
event-free survival distribution follows directly from
$p_y(t,\bm{x})$ as in \cite{SparLoga16} using the expression
\[
S(t_{(j)}|\bm{x})=\prod_{l=1}^j (1-p_y(t_{(l)},\bm{x})), j=1,\ldots,k.
\]
Inference on the cumulative incidence for cause~1 can be carried out
using the expression
\[
F_1(t_{(j)}|\bm{x})=\sum_{l=1}^j S(t_{(l-1)}|\bm{x})\psi(t_{(l)},\bm{x}).
\]
With these functions in hand, one can easily accomplish inference for
other quantities of interest based on the cumulative incidence
function, such as conditional quantiles \cite{PengFine09} defined as
$Q_1(\tau|\bm{x})=\inf\{t:F_k(t|\bm{x} \geq \tau)\}$.  Analogous
expressions for the cumulative incidence for the competing causes are
also directly available.  Note that Method~1 can easily be extended to
multiple causes, i.e., cause~1 vs.\ cause~2 vs.\ cause~3, etc.

\subsection{Method 2}
In this method, we define $\tilde{p}_{ij2}=p_{ij2}/(1-p_{ij1})$ as the
conditional probability of event 2 at time $t_{(j)}$ for patient $i$
given that no cause 1 event occurred at that time, and re-express the
likelihood as follows.

\begin{eqnarray}\label{eq:survplusbinary}
L(p|\bm{y})\ &=& \prod_{i=1}^N\prod_{j=1}^{n_i} p_{ij1}^{y_{ij1}}[\tilde{p}_{ij2}(1-p_{ij1})]^{y_{ij2}}(1-p_{ij1}-\tilde{p}_{ij2}(1-p_{ij1}))^{1-y_{ij1}-y_{ij2}} \notag \\
&=& \prod_{i=1}^N\prod_{j=1}^{n_i} p_{ij1}^{y_{ij1}} (1-p_{ij1})^{1-y_{ij1}} \prod_{i=1}^N\prod_{j:y_{ij1}=0} \tilde{p}_{ij2}^{y_{ij2}}(1-\tilde{p}_{ij2})^{1-y_{ij2}} .
\end{eqnarray}

This likelihood also separates into two binary likelihoods, so that
we can fit separate BART probit models for $p_{ij1}$ and
$\tilde{p}_{ij2}$, using the corresponding binary observations
$y_{ij1}$ and $y_{ij2}$ respectively.  Specifically, we assume
\begin{equation}\label{disp:bartmodel1}
\begin{array}{rcl}
y_{ij1} & = & I_{z_{ij1} \ge 0} \\
z_{ij1} & = & \mu_{1} + f_{1}(t_{(j)},\bm{x}_{i}) + e_{ij1} \\
f_{1} & \sim & \bart{m, \mu_{1}, \tau, \alpha, \gamma}
\end{array}
\end{equation}
for the first model and 
\begin{equation}\label{disp:bartmodel1}
\begin{array}{rcl}
y_{ij2} & = & I_{z_{ij2} \ge 0} \\
z_{ij2} & = & \mu_{2} + f_2(t_{(j)},\bm{x}_{i}) + e_{ij2} \\
f_{2} & \sim & \bart{m, \mu_{2}, \tau, \alpha, \gamma}
\end{array}
\end{equation}
for the second model.  Conceptually, the first BART function models
the conditional probability of a cause 1 event at time $t_{(j)}$, given
the patient is still at risk prior to time $t_{(j)}$, while the second
BART function models the conditional probability of a cause 2 event at
time $t_{(j)}$, given the patient is still at risk prior to time
$t_{(j)}$ and does not experience a cause 1 event.  As above, the
algorithms in existing BART software provide for samples from the
posterior distribution of $f_1$ and $f_2$ given the data.  Similarly,
samples from the posterior distribution of
$p_1(t,\bm{x})=\Phi(\mu_{1}+f_{1}(t,\bm{x}))$ and
$p_2(t,\bm{x})=\Phi(\mu_{2}+f_{2}(t,\bm{x}))$ can be obtained.
Samples from the event-free survival distribution are obtained from
the expression
\[
S(t_{(j)}|\bm{x})=\prod_{l=1}^j (1-p_1(t_{(l)},\bm{x}))(1-p_2(t_{(l)},\bm{x})), j=1,\ldots,k.
\]
Samples from the cumulative incidence for cause~1 can be obtained
using the expression
\[
F_1(t_{(j)}|\bm{x})=\sum_{l=1}^j S(t_{(l-1)}|\bm{x})p_1(t_{(l)},\bm{x}).
\]
\subsection{Data construction}

Competing risks data contained in observations $(t,\delta,\epsilon)$
must be recast as binary outcome data; similarly, the corresponding
time variable is recast as a covariate in order to fit the BART
models described in both methods above.  For additional clarification,
we give a very simple example of a data set with three
observations here:
\[
(t_1,\delta_1,\delta_1\epsilon_1)=(2.5,1,1),\ (t_2,\delta_2,\delta_2\epsilon_2)=(1.5,1,2),\ (t_3,\delta_3,\delta_3\epsilon_3)=(3,0,0)\] 
where $t_{(1)}=1.5,\ t_{(2)}=2.5,\ t_{(3)}=3$.

For observation 1, the patient is at risk at time $t_{(1)}=1.5$, but
does not experience an event, so that
$y_{111}=0,y_{112}=0,y_{11\cdot}=0$.  This same patient experiences a
cause 1 event at time $t_{(2)}=2.5$, so that $y_{121}=1$ and
$y_{12\cdot}=1$.  However, because they experienced a cause 1 event at
time $t_{(2)}$, the patient is no longer at risk of experiencing a
cause 2 event using the Method~2 formulation of conditional
probabilities, so we do not include $y_{122}$.  For observation 1,
$u_1=1$ since the patient experienced a cause 1 event at time
$t_1=t_{(2)}$.  For observation 2, since the patient experiences a
cause 2 event at time $t_{(1)}=1.5$, we define $y_{211}=0$,
$y_{212}=1$, $y_{21\cdot}=1$, and $u_2=0$.  For observation 3, since
the patient is censored at $t_{(3)}=3$, all $y_{3jk}=0$ for
$j=1,\ldots,3$ and $k=1,2,\cdot$.  Also, there is no $u_3$ defined for
this patient since they did not experience any kind of event.  A summary of
the binary indicators and corresponding time covariates for each
binary observation are summarized in Table~\ref{dataconstruct}.
Besides time, the remaining covariates would contain the individual
level covariates, $\bm{x}_i$, with rows repeated to match the
repetition pattern of the first subscript of $\bm{y}$.

\begin{table}
\small\sf\centering
\caption{Data Construction Example \label{dataconstruct}}
\begin{tabular}{ccccccc}
%\toprule
 & & &\multicolumn{2}{c}{Method 1}&\multicolumn{2}{c}{Method 2} \\
$i$ &$j$ &$t_{(j)}$& $y_{ij\cdot}$ &$u_i$&$y_{ij1}$ &$y_{ij2}$ \\
%\midrule
1&1&1.5&0& &0&0\\
 &2&2.5&1&1&1&  \\
2&1&1.5&1&0&0&1 \\
3&1&1.5&0& &0&0 \\
 &2&2.5&0& &0&0 \\
 &3&3.0&0& &0&0 \\
%\bottomrule
\end{tabular}
\end{table}

\section{Performance of proposed methods}

In order to determine the operating characteristics of our new
method, we conducted several simulation studies and summarized various prediction performance metrics.  We start with a two-sample setting to establish the face validity of the method to handle competing risks data with two groups.   We then move on to examine performance in a complex regression setting.  

\subsection{Two sample setting}

With a two sample scenario, several settings are considered to
represent standard modeling approaches to competing risks data: 1)
proportional cause-specific hazards data generated from a Cox model;
2) proportional subdistribution hazards data generated from a Fine and
Gray model; and 3) nonproportional subdistribution setting based on
Weibull distributions.  In each case, we simulate data sets with
sample sizes of $N=250, 500, 1000$ under independent exponential
censoring with rate parameters leading to overall censoring
proportions of $20\%$ or $50\%$.  Four different parameter settings
are considered for each case. A total of 400 replicate data sets were
generated in each instance.

\newpage

{\bf Case 1: Proportional cause-specific hazards generated by Cox model}

For $x \in \{0,1\}$ and failure cause $k \in \{0,1\}$, the cause
specific hazard is given by
$\lambda_k(t, x) = \lambda_{0k} \e{x \beta_k} \where \lambda_{0k}>0 $.
The cumulative hazard for any cause of failure is given by
$ \Lambda(t, x) =(\lambda_{01}\e{x \beta_1} + \lambda_{02}\e{x
  \beta_2})t $, and the cumulative incidence for cause $k$ is given by
\[
  F_k(t, x)  =  \frac{\lambda_k(t, x)}{\lambda_1(t, x)+\lambda_2(t, x)} \left[ 1-\e{-\Lambda(t, x)}\right] .\] 
The limiting cumulative incidence for cause $1$ in group $x$ is 
\[
  p_x=F_1(\infty, x) = \frac{\lambda_{01}\e{x\beta_1}}{\lambda_{01}\e{x\beta_1}+\lambda_{02}\e{x\beta_2}} .\]

  {\bf Case 2: Proportional subdistribution hazards generated by Fine
    and Gray model}

  Under a proportional subdistribution hazards model
  \citep{FineGray99}, the cumulative incidence functions
  can be directly specified as in \citep{LogaZhan13} as
\begin{align}
\label{fgcif}
  F_1(t, x) & = 1-\wrap{1-p_0 (1-\e{-\gamma_0 t})}^{\e{x \beta_1}}  \\
  F_2(t, x) & = (1-p_0)^{\e{x \beta_1}} (1-\e{-\gamma_0 t}) 
\end{align}

{\bf Case 3: Nonproportional hazards based on Weibull subdistributions}

To simulate this scenario, we describe a data generation process where
first the failure cause is generated with probability $p_0$ for cause
1 regardless of group, and conditional on the failure cause, the
failure time is generated from a Weibull distribution with scale
parameter $\gamma_0$ and shape parameter $e^{x\beta_k}$.  Because the
shape parameter is group dependent, this leads to different shapes of
the cumulative incidence functions, with the same limiting cumulative
incidence.  The resulting cumulative incidence functions have the
following form:
\[ F_k(t,x)=p_0^{2-k}(1-p_0)^{k-1} (1-e^{-\gamma_0t^{e^x\beta_k}}).\]
%\[ F_k(t,x)=p_0^{2-k}(1-p_0)^{k-1} (1-e^{-\gamma_0t^{e^x\beta_k}}).\]

A summary of the parameter settings studied are in Table~\ref{cases} below.  

\begin{table}
\caption{Parameter settings for Cases 1 through 3. \label{cases}}
\begin{tabular}{lllrrllll}
Case & $\lambda_{01}$ & $\lambda_{02}$ & $\beta_1$ & $\beta_2$ & $p_0$ & $p_1$
 & $\gamma_0$ & $\gamma_1$  \\ \hline
1,               & 1   & 1   & 0         & 0        & 0.5 & 0.5 &    &   \\
Proportional     & 1   & 1   & $-\log 2$ & $\log 2$ & 0.5 & 0.2 &    & 2.5  \\
Cox              & 2   & 0.5 & 0         & 0        & 0.8 & 0.8 &  &   \\
                 & 2   & 0.5 & $-\log 2$ & $\log 2$ & 0.8 & 0.5 &  &   \\ \hline
2,               &  &  & 0         &          & 0.5 &     & 2   &  \\
Subdistribution  &  &  & $-\log 2$ &          & 0.5 &     & 2   &  \\
Fine and Gray    &  &  & 0         &          & 0.8 &     & 2.5 &  \\
                 &  &  & $ \log 2$ &          & 0.2 &     & 2.5 &   \\ \hline
3,               &  &  & 0         & 0        & 0.5 &     & 2   & \\
Nonproportional  &  &  & $-\log 3$ & $\log 3$ & 0.5 &     & 2   &  \\
Weibull-like     &  &  & 0         & 0        & 0.8 &     & 2.5 &  \\
                 &  &  & $-\log 3$ & $\log 3$ & 0.2 &     & 2.5 &  \\ \hline
\end{tabular}
\end{table}

Each simulated data set was analyzed with both BART competing risks
models, Cox proportional cause specific hazards models \citep{Cox72},
Fine and Gray proportional subdistribution hazards model
\citep{FineGray99}, and the Aalen-Johansen nonparametric estimator
\citep{AaleJoha78} applied separately to each group.  For brevity, we
only consider cause~1 which is generally the cause of interest.  For
each scenario, we examined the prediction performance in terms of Root
Mean Square Error (RMSE) and bias, at the following quantiles of the
event-free survival (with either failure cause as an event)
distribution: 10\%, 30\%, 50\%, 70\% and 90\%.  We also compare the
95\% interval coverage probability and 95\% interval length for the
two BART methods.  Results are plotted as points against quantile
for each case and sample combination; note that there are 16 points
for each case and sample combination, representing 2 censoring
percentages, 4 parameter configurations, and 2 groups as targets for
prediction.

\begin{figure}  
\begin{minipage}[b]{.5\linewidth} \centering\large 
\includegraphics[scale=0.35]{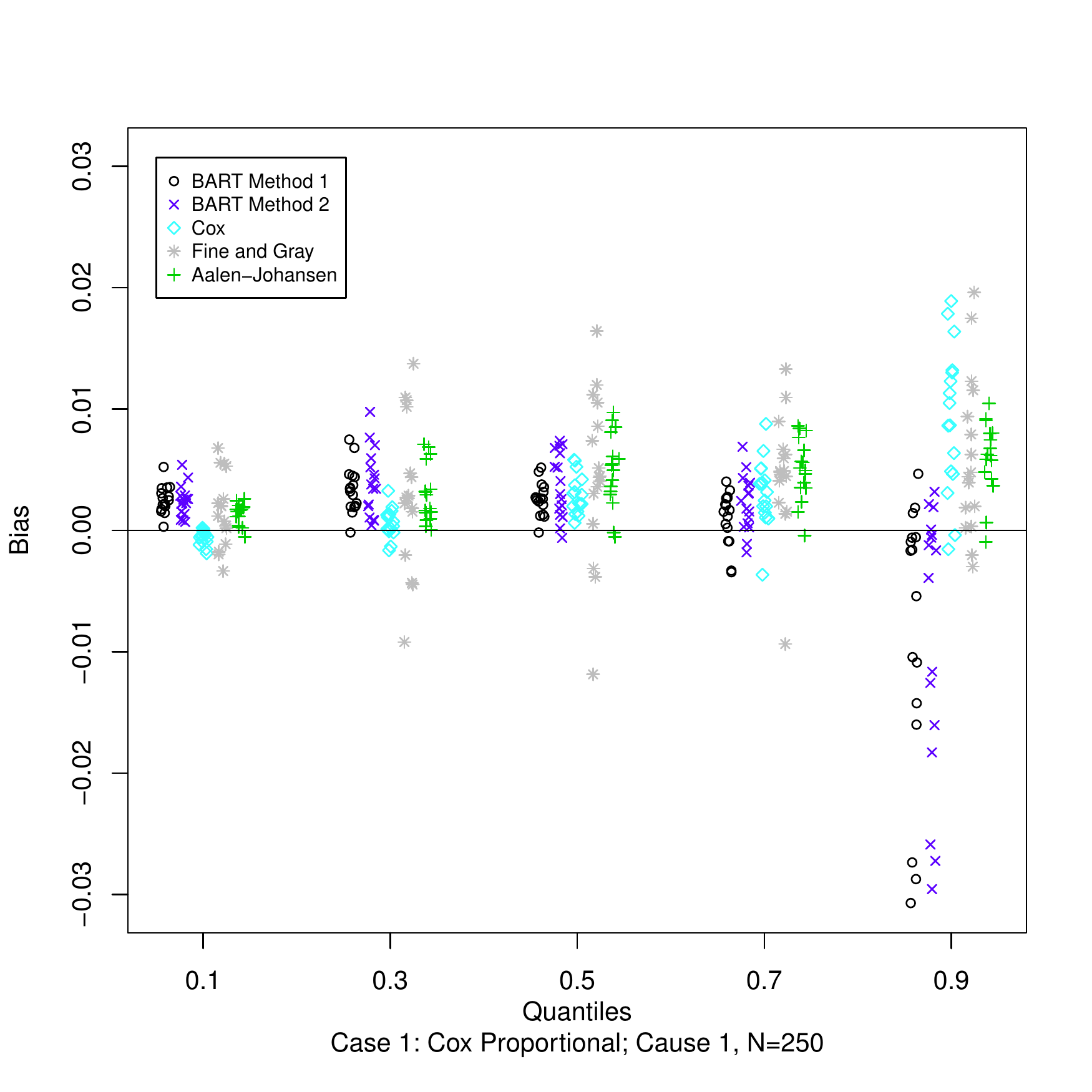}
 \label{fig:1a} \\
\includegraphics[scale=0.35]{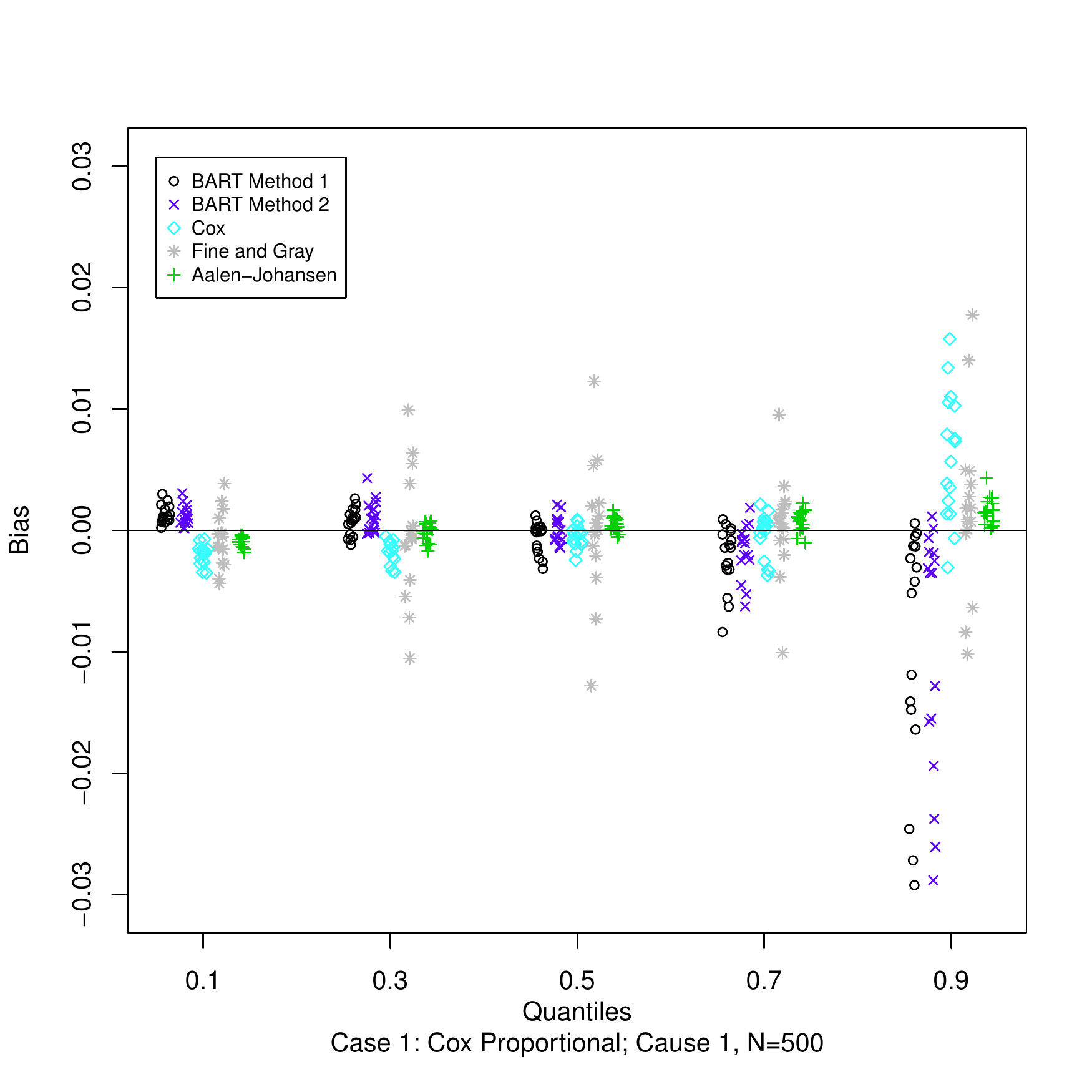}
 \label{fig:1c}  \\
\includegraphics[scale=0.35]{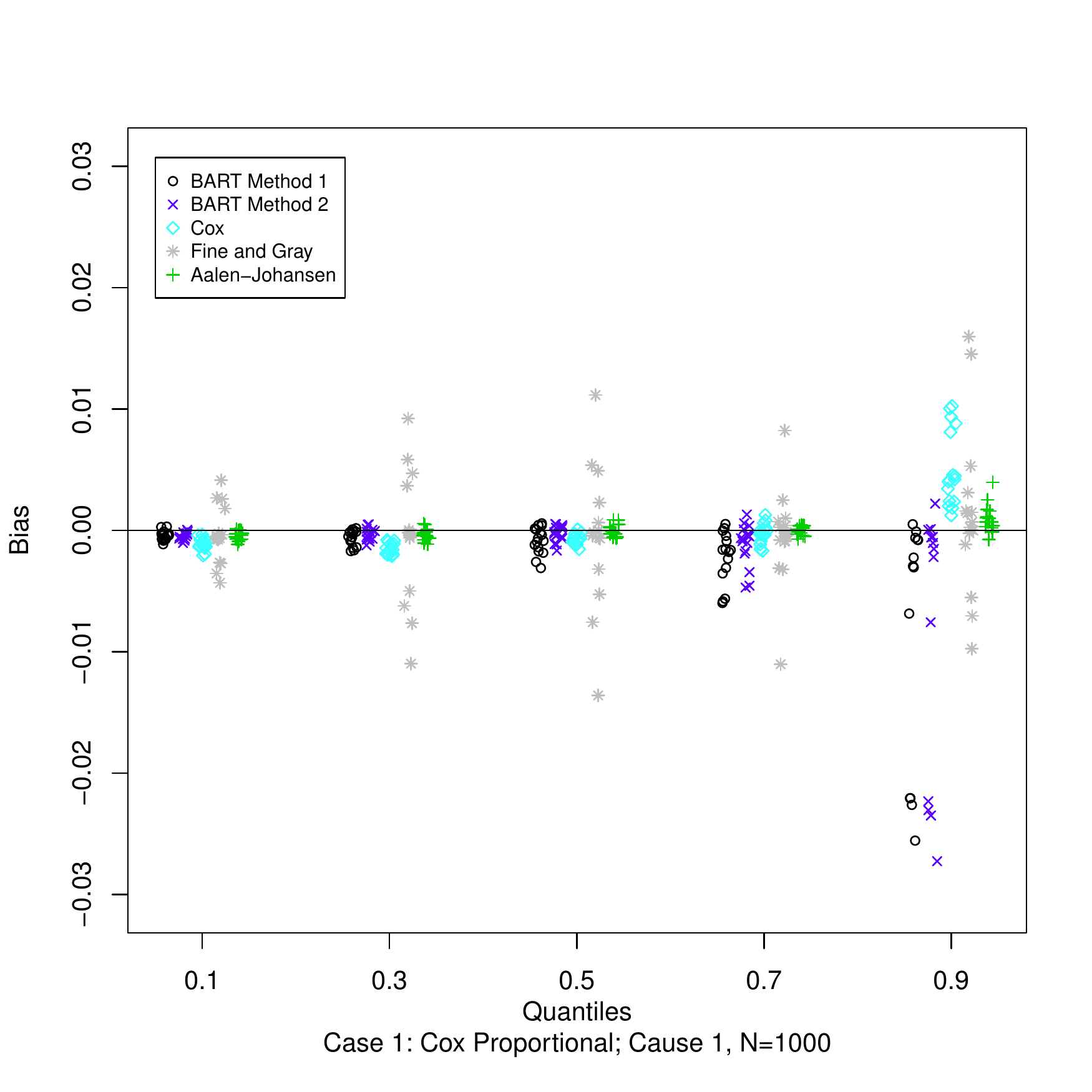}
 \label{fig:1e}
\end{minipage} 
\begin{minipage}[b]{.5\linewidth} \centering\large 
\includegraphics[scale=0.35]{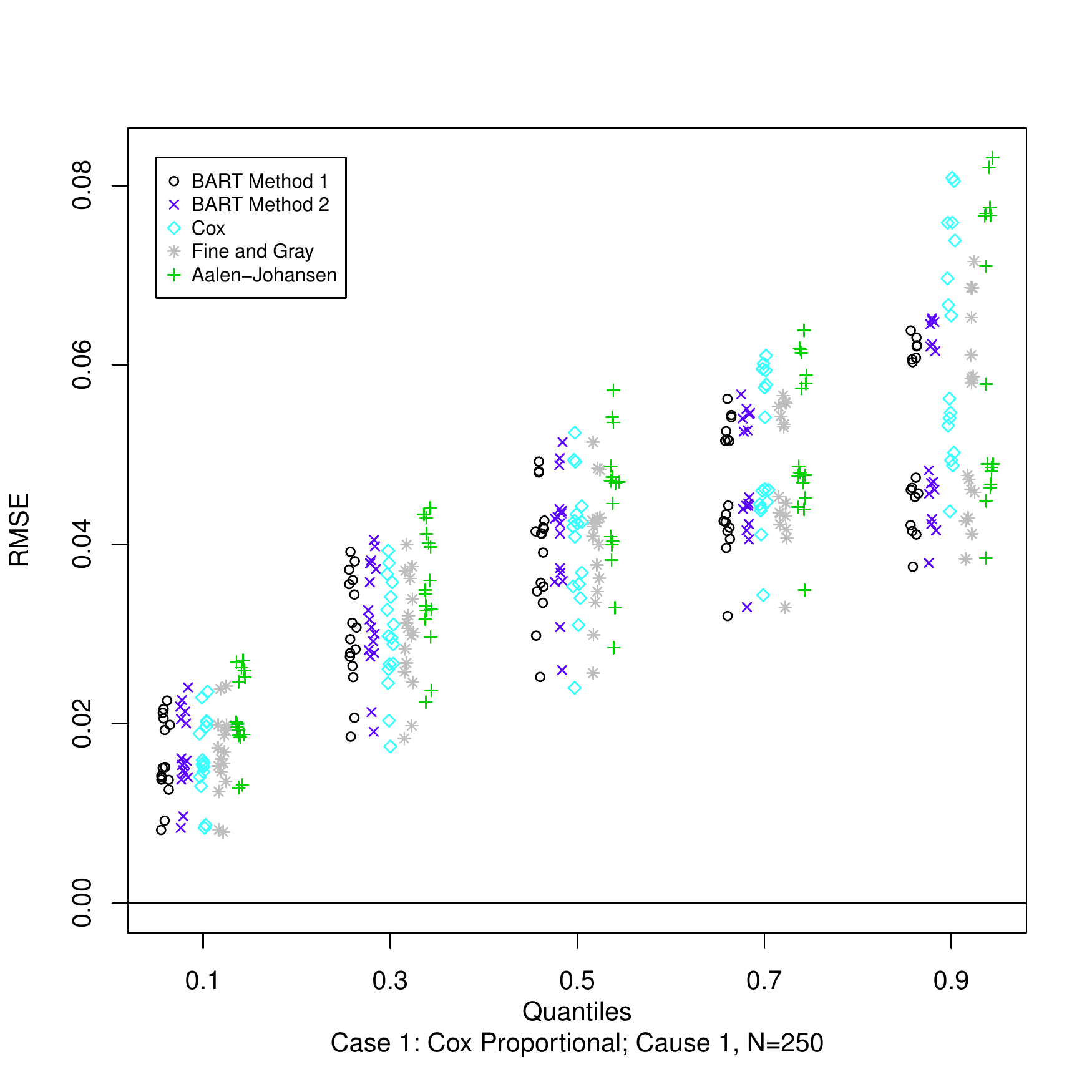} 
 \label{fig:1b} \\
\includegraphics[scale=0.35]{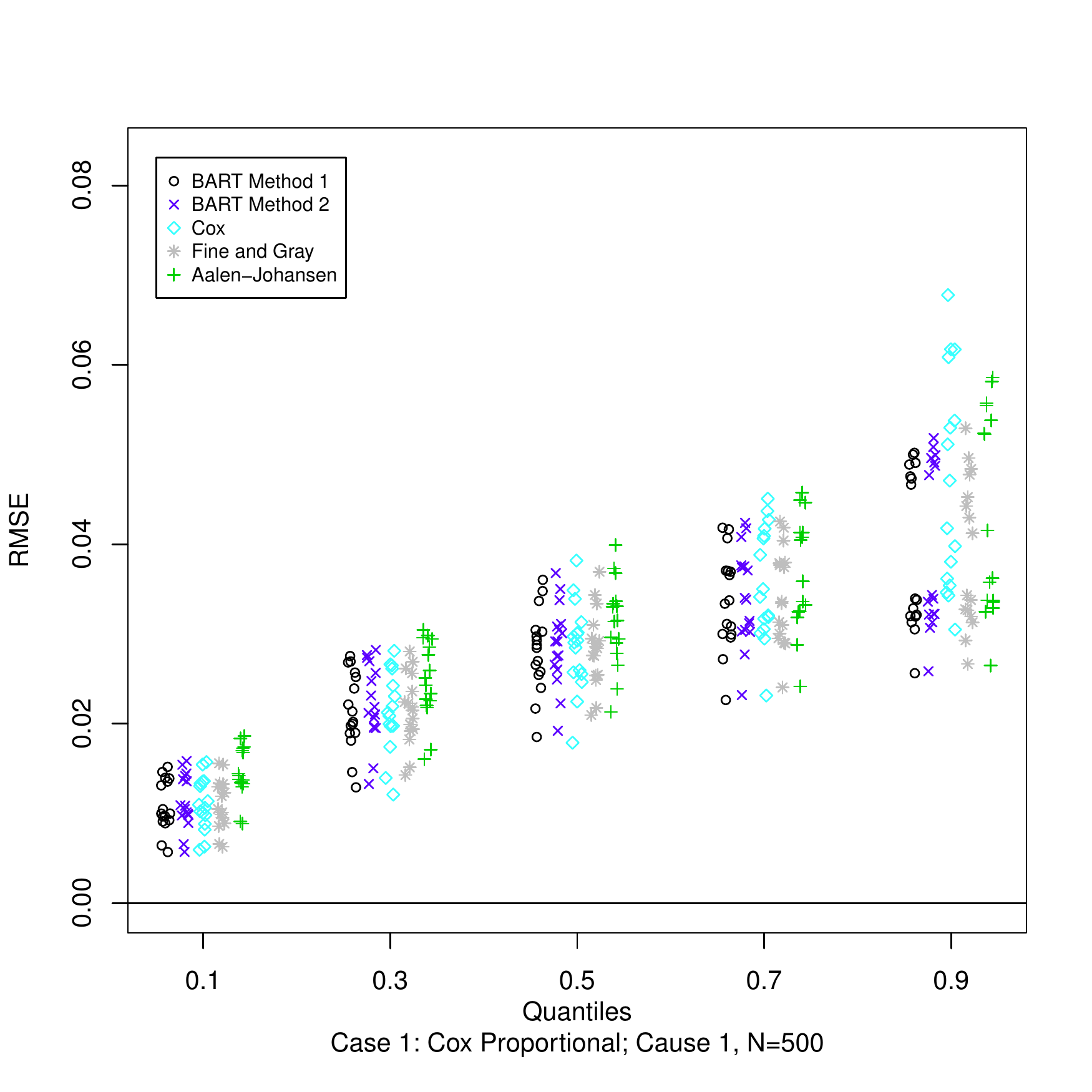}
\label{fig:1d} \\
\includegraphics[scale=0.35]{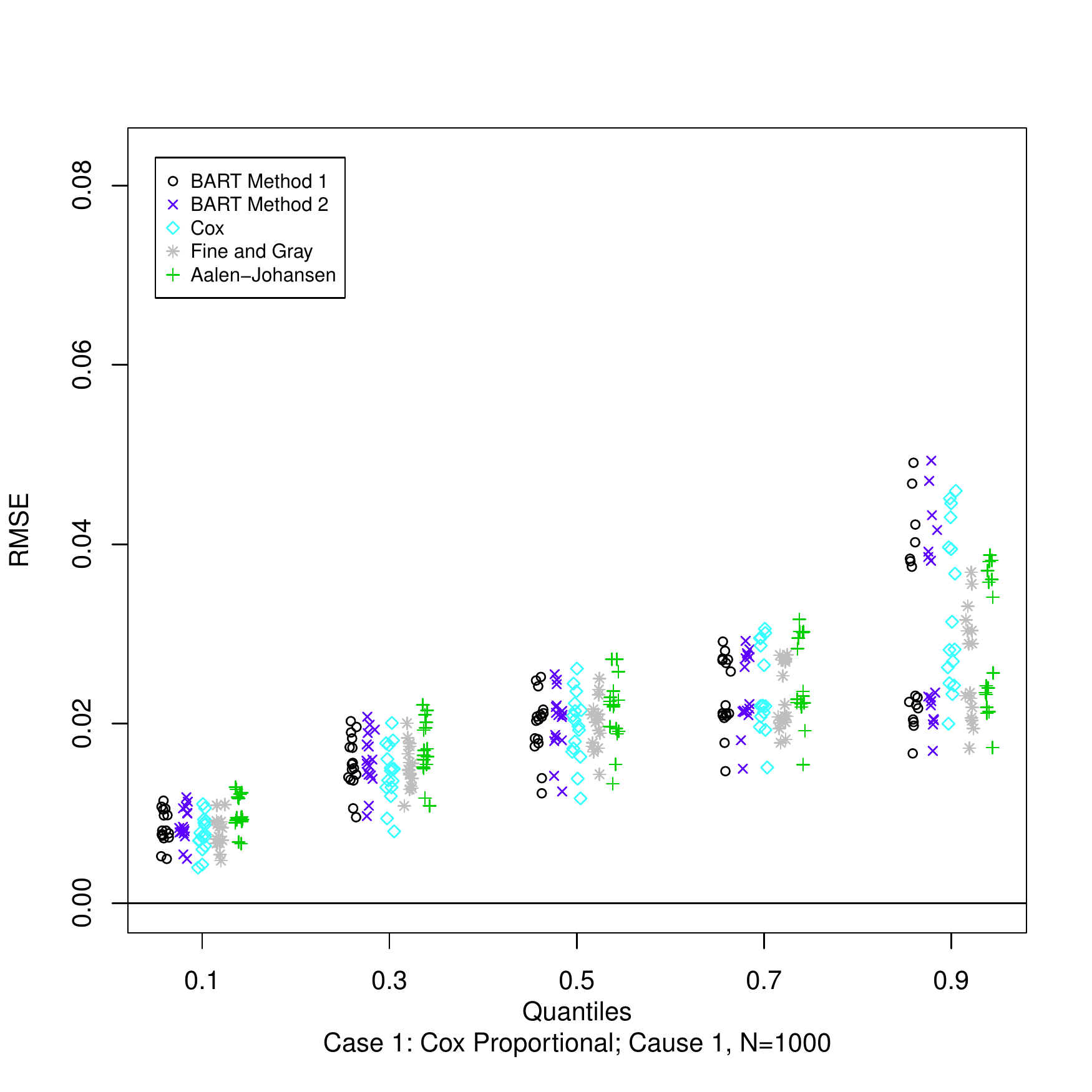}
\label{fig:1f} 
\end{minipage}% 
\caption{Bias (left) and RMSE (right) for case 1, $N=250$ (first row), $N=500$ (second row), and $N=1000$ (third row).  }\label{fig:1} \end{figure}

\begin{figure}  
\begin{minipage}[b]{.5\linewidth} \centering\large 
\includegraphics[scale=0.35]{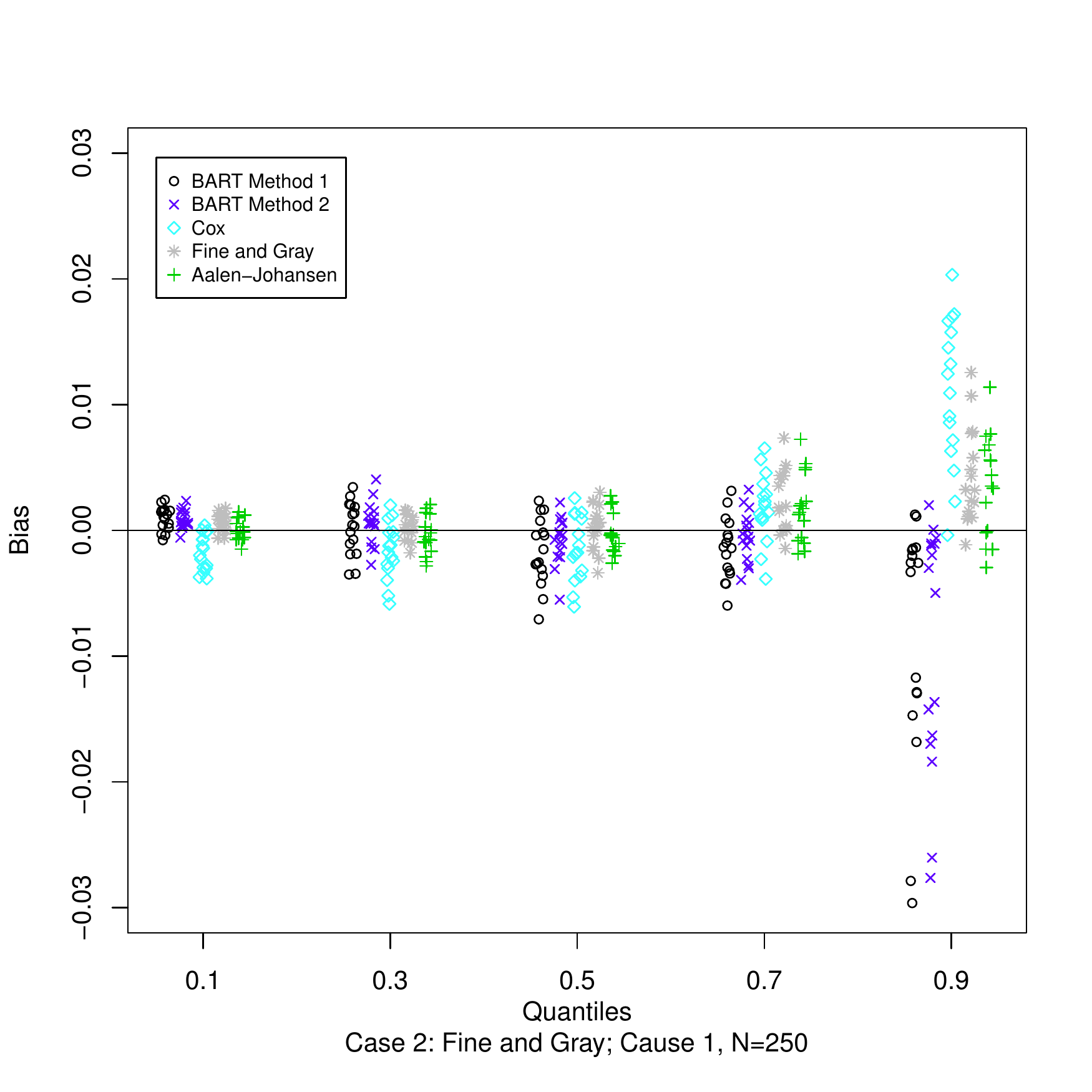}
 \label{fig:2a} \\
\includegraphics[scale=0.35]{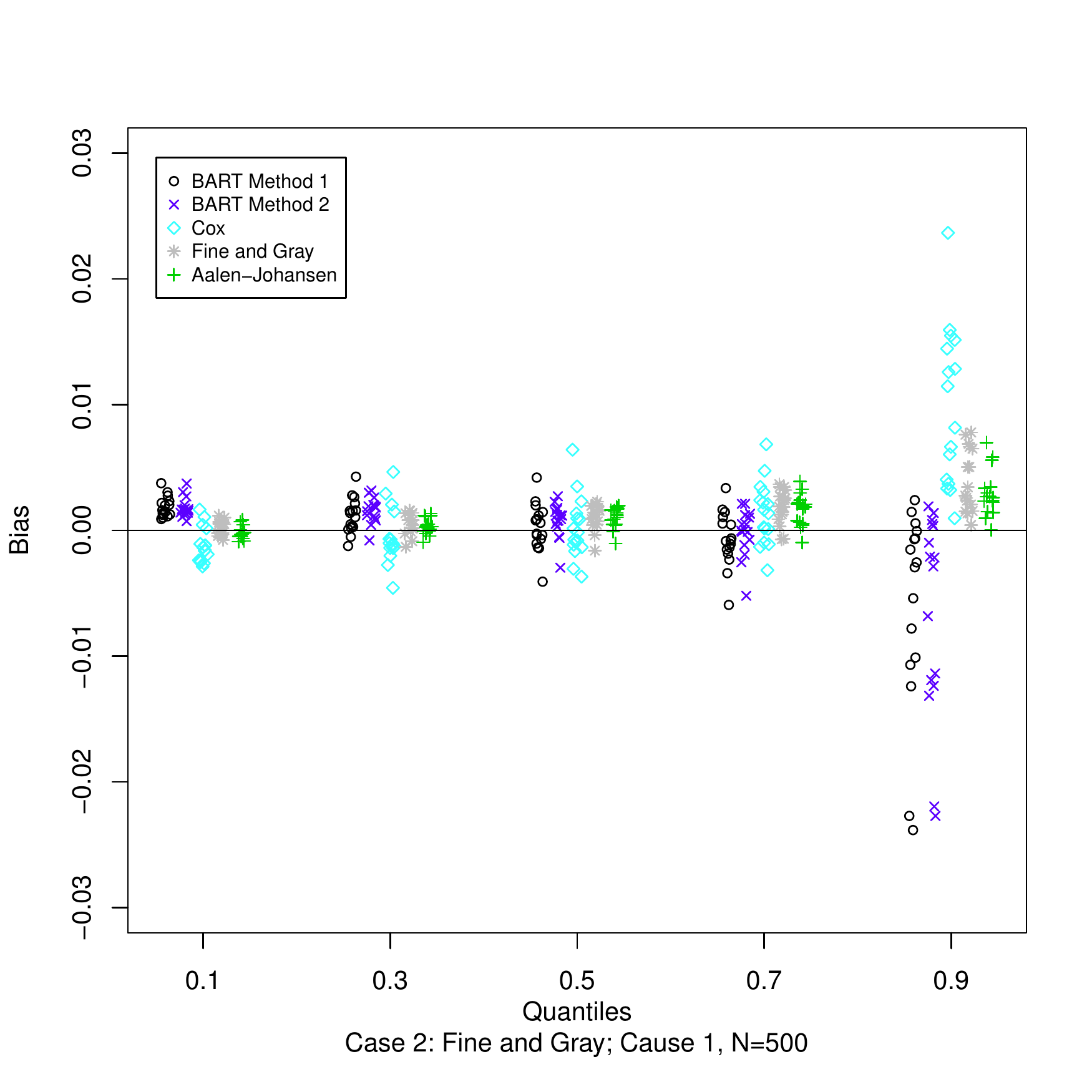}
 \label{fig:2c}  \\
\includegraphics[scale=0.35]{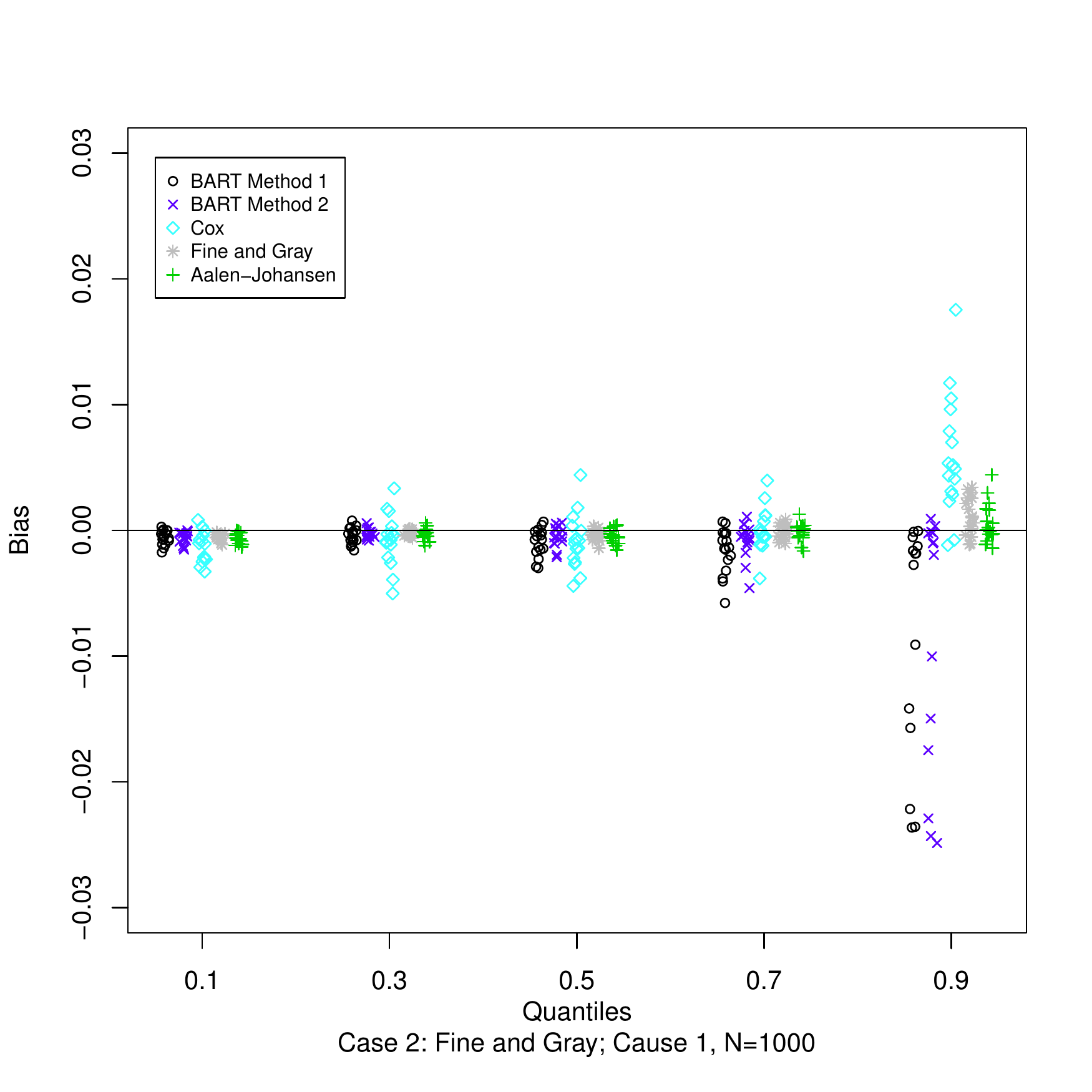}
 \label{fig:2e}
\end{minipage} 
\begin{minipage}[b]{.5\linewidth} \centering\large 
\includegraphics[scale=0.35]{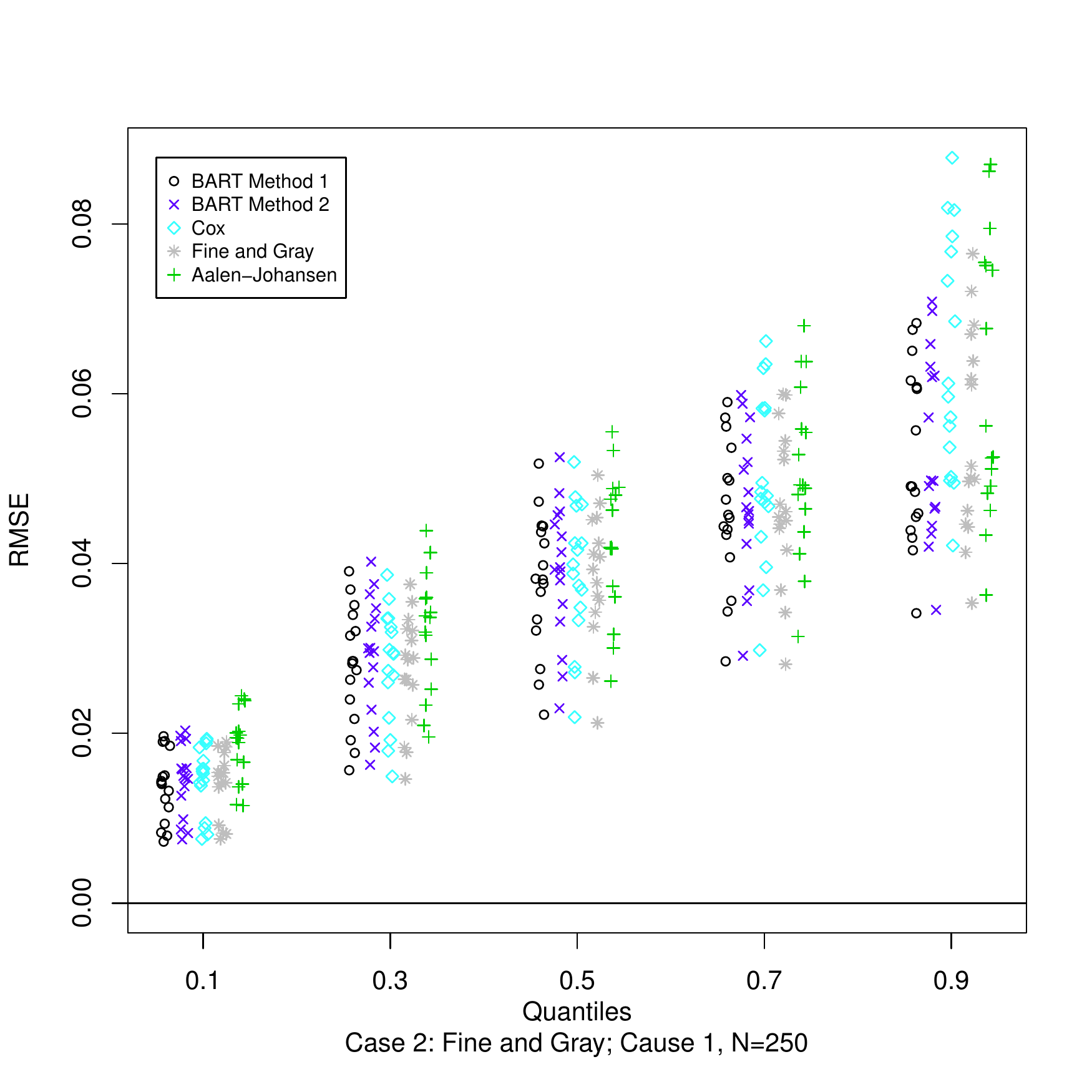} 
 \label{fig:2b} \\
\includegraphics[scale=0.35]{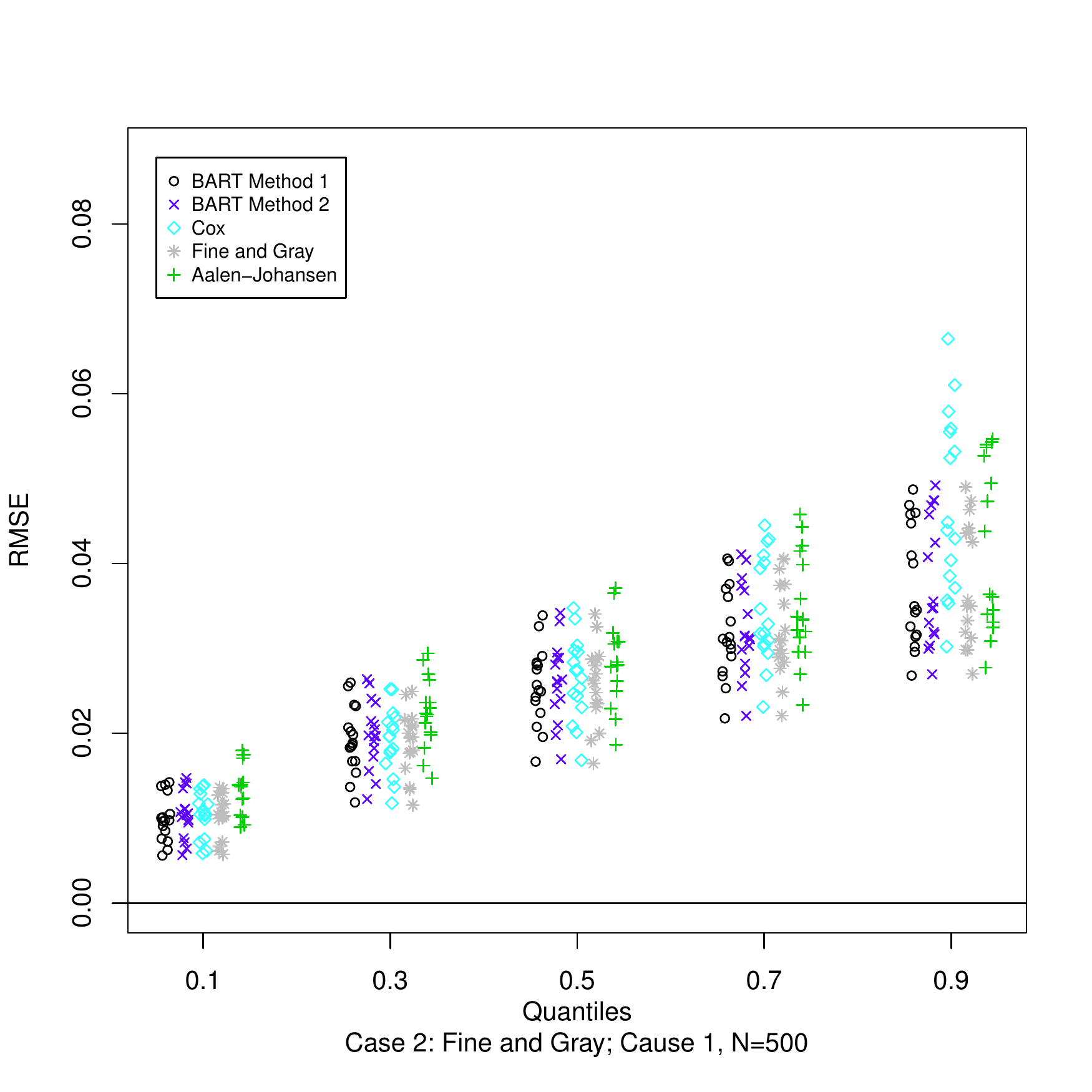}
\label{fig:2d} \\
\includegraphics[scale=0.35]{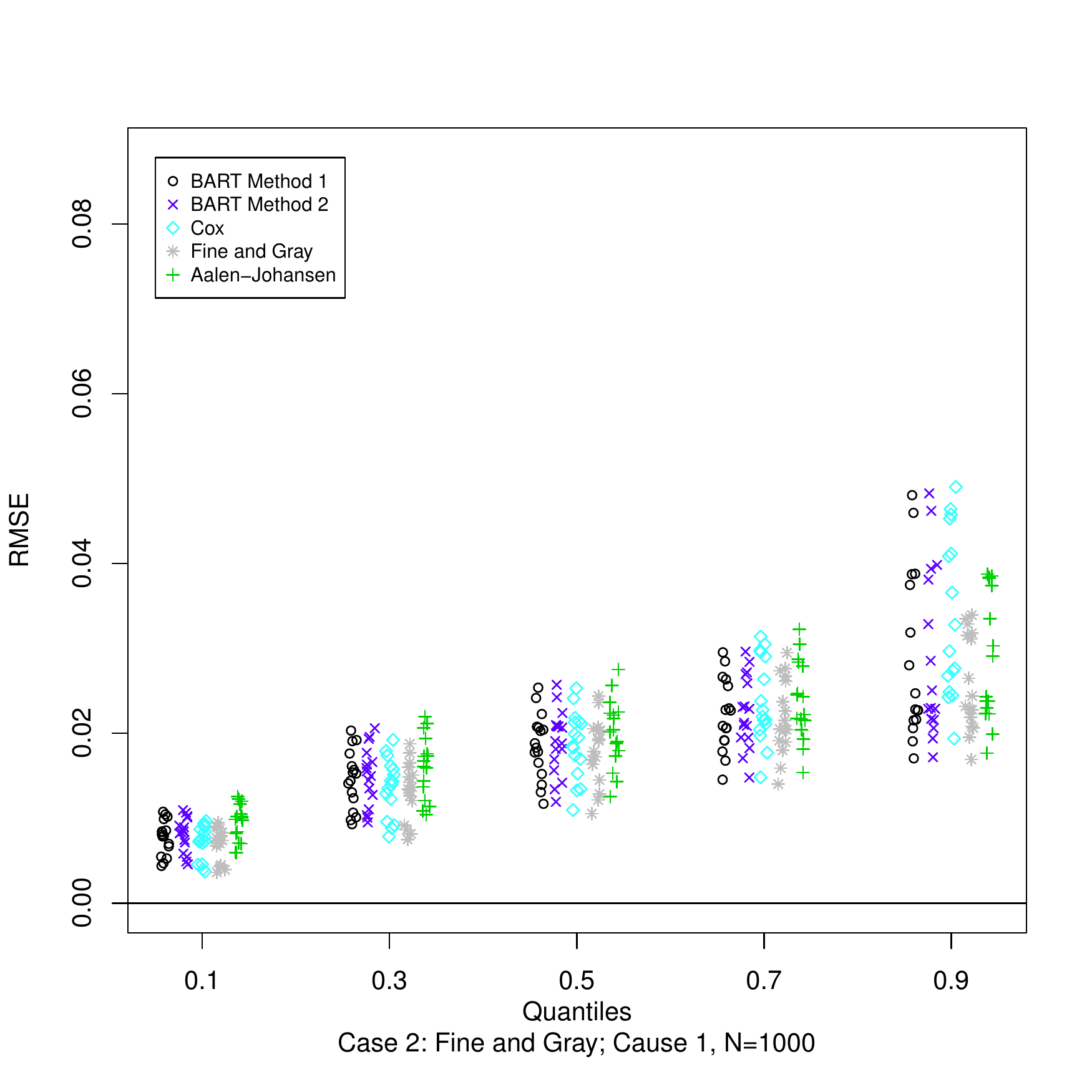}
\label{fig:2f} 
\end{minipage}% 
\caption{Bias (left) and RMSE (right) for case 2, $N=250$ (first row), $N=500$ (second row), and $N=1000$ (third row).  }\label{fig:2} \end{figure}

\begin{figure}  
\begin{minipage}[b]{.5\linewidth} \centering\large 
\includegraphics[scale=0.35]{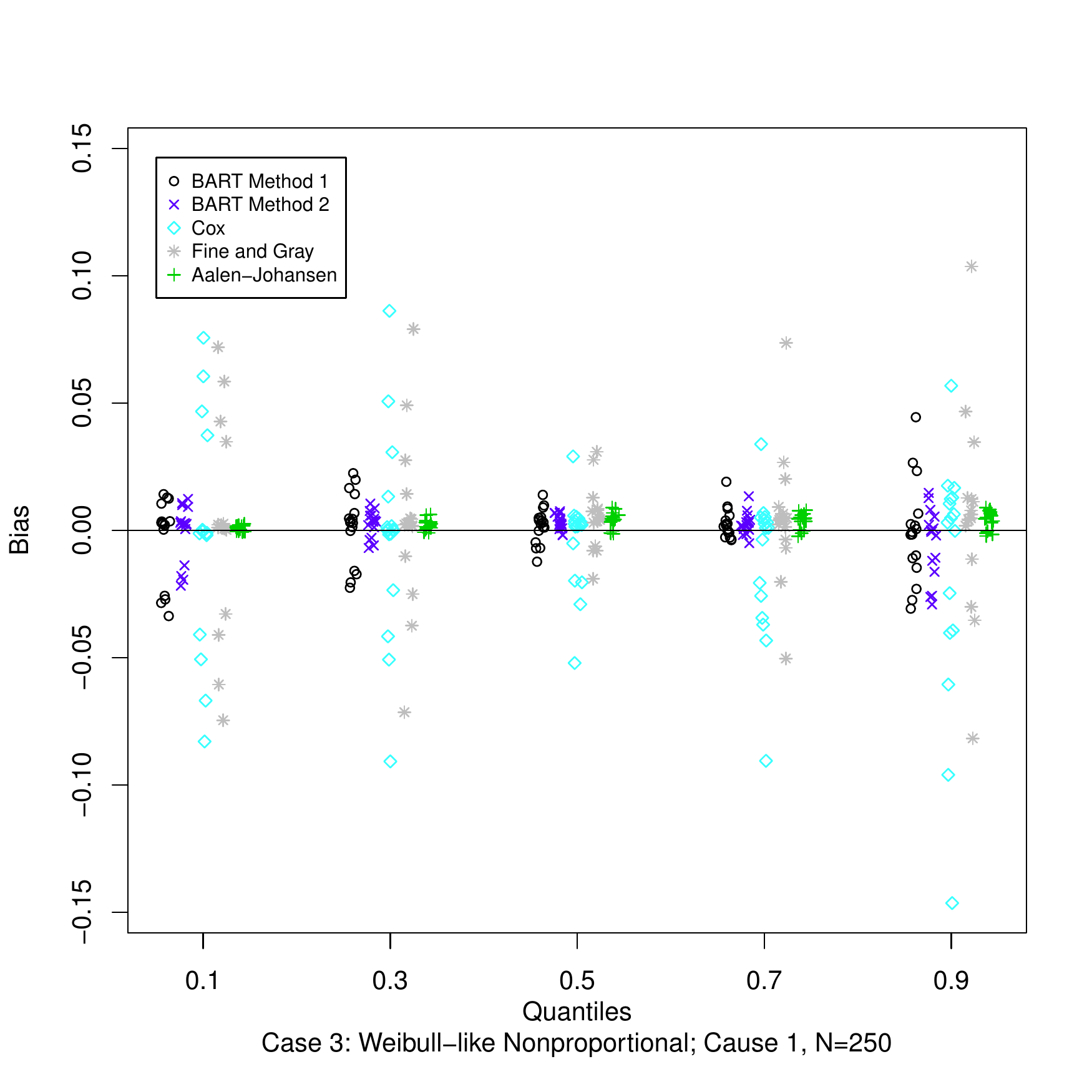}
 \label{fig:3a} \\
\includegraphics[scale=0.35]{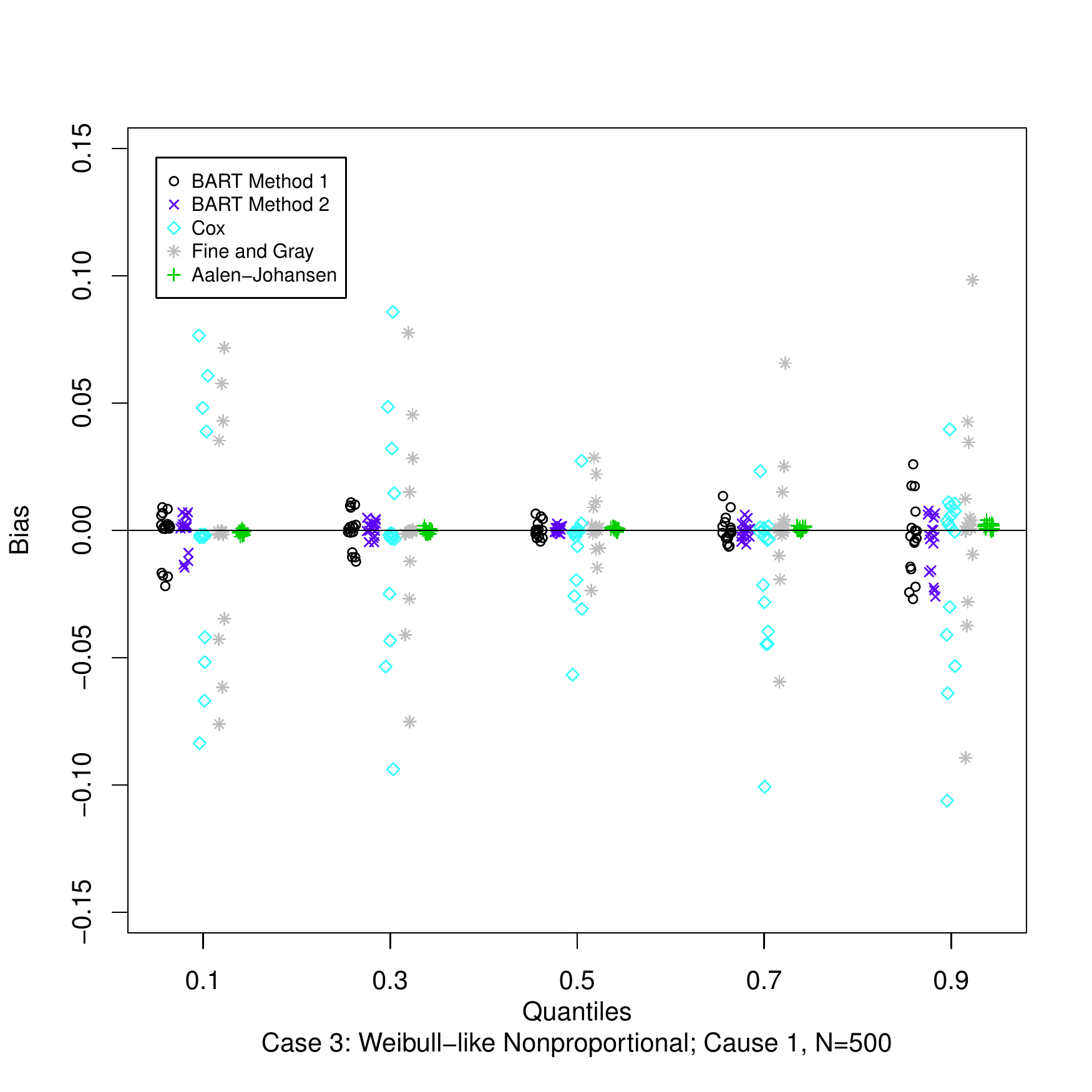}
 \label{fig:3c}  \\
\includegraphics[scale=0.35]{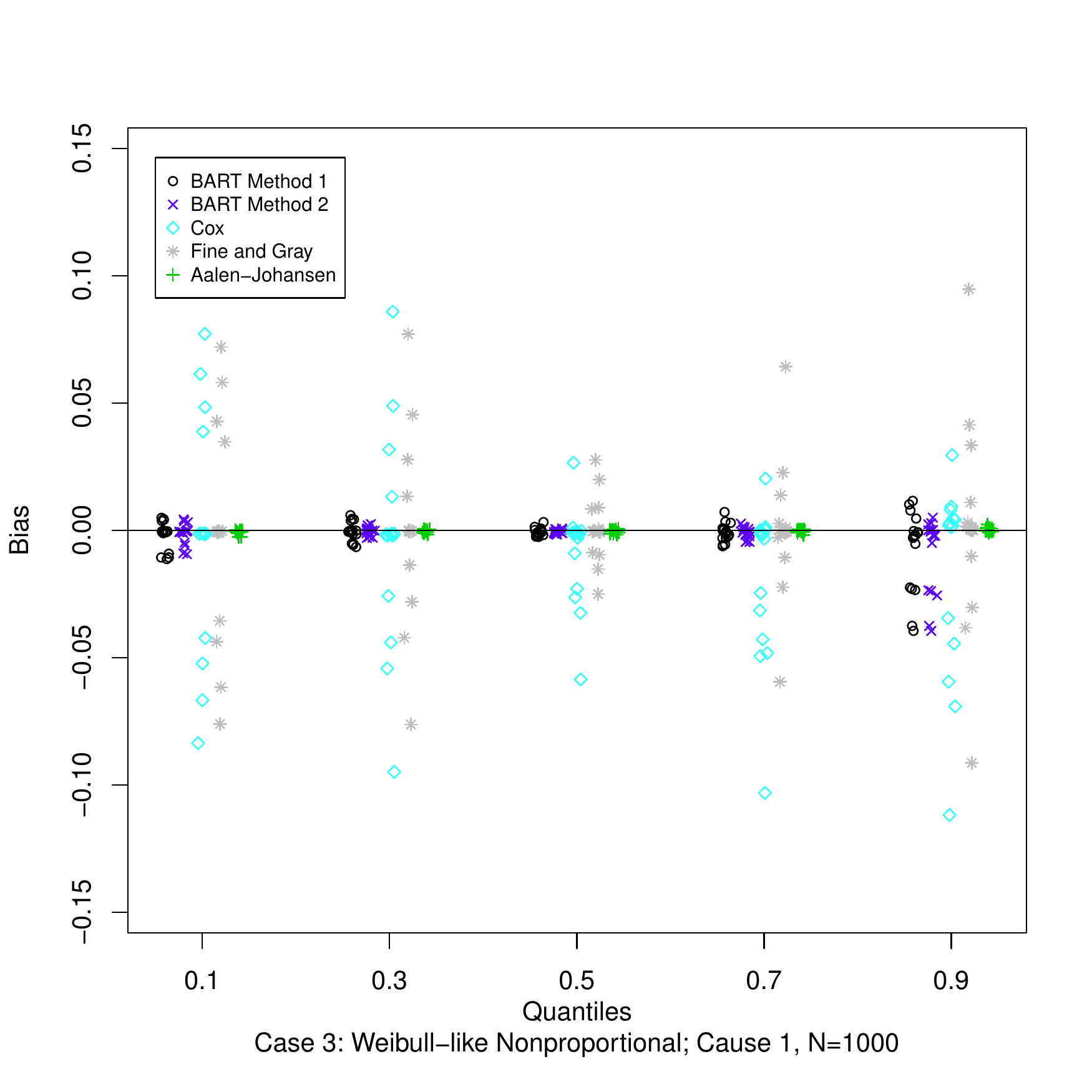}
 \label{fig:3e}
\end{minipage} 
\begin{minipage}[b]{.5\linewidth} \centering\large 
\includegraphics[scale=0.35]{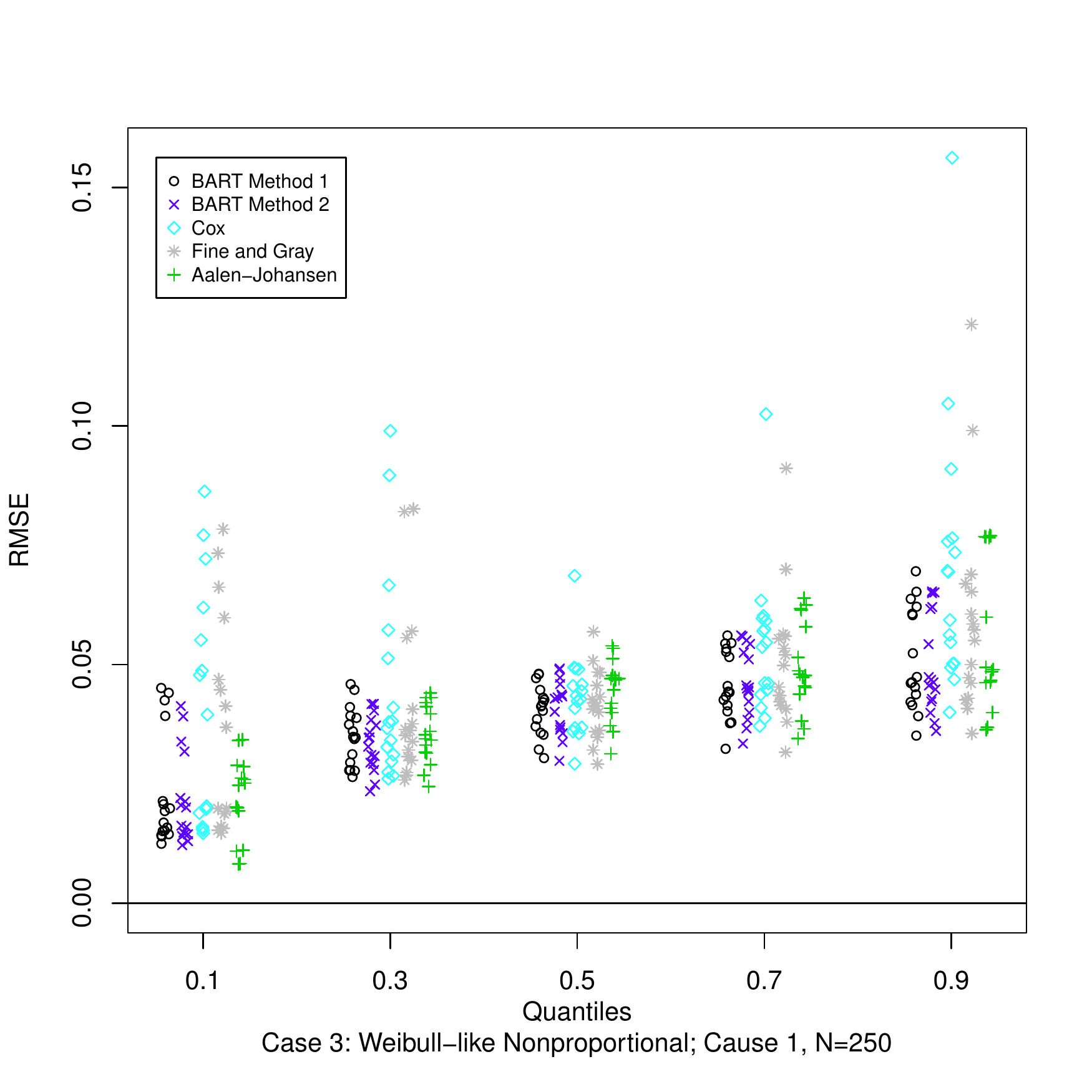} 
 \label{fig:3b} \\
\includegraphics[scale=0.35]{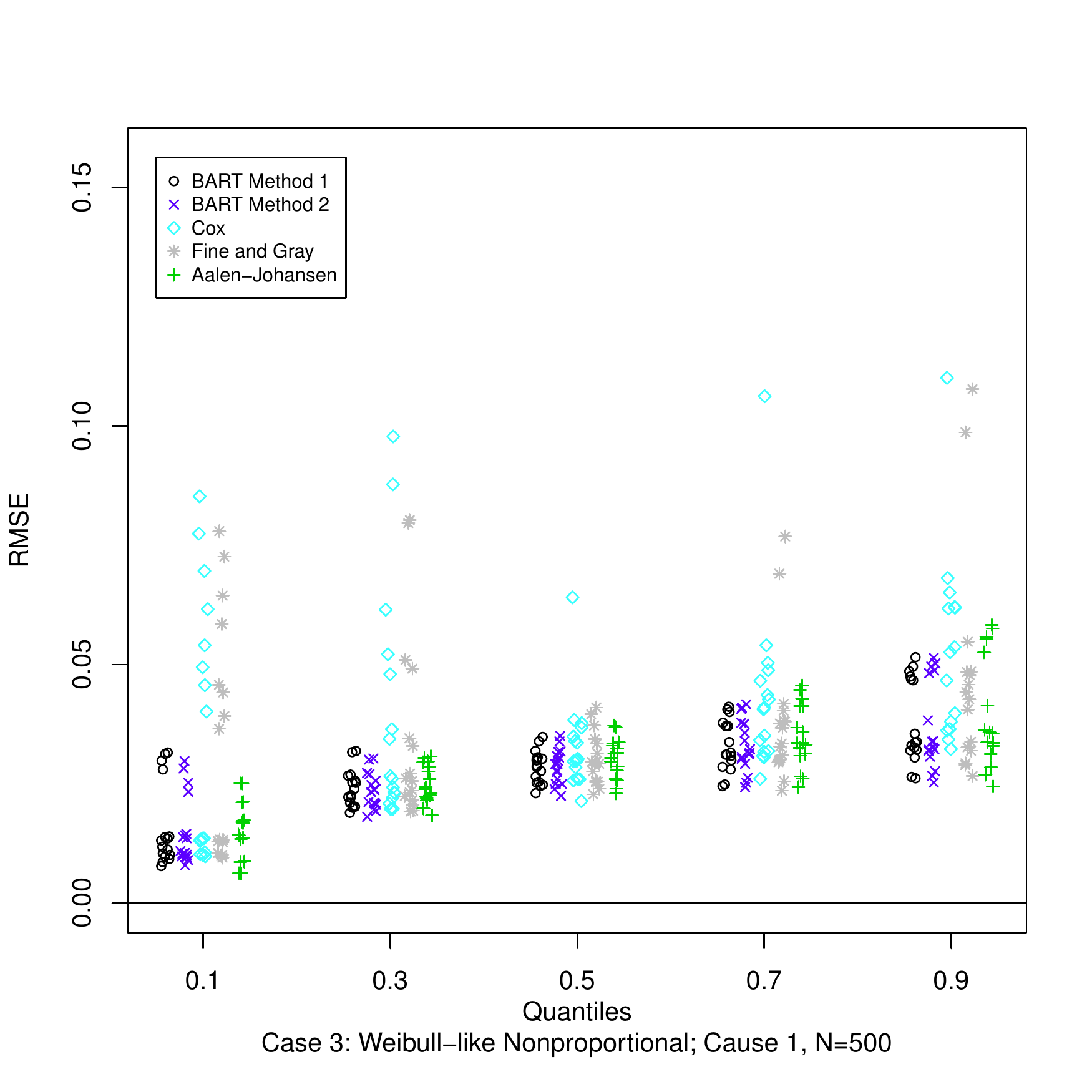}
\label{fig:3d} \\
\includegraphics[scale=0.35]{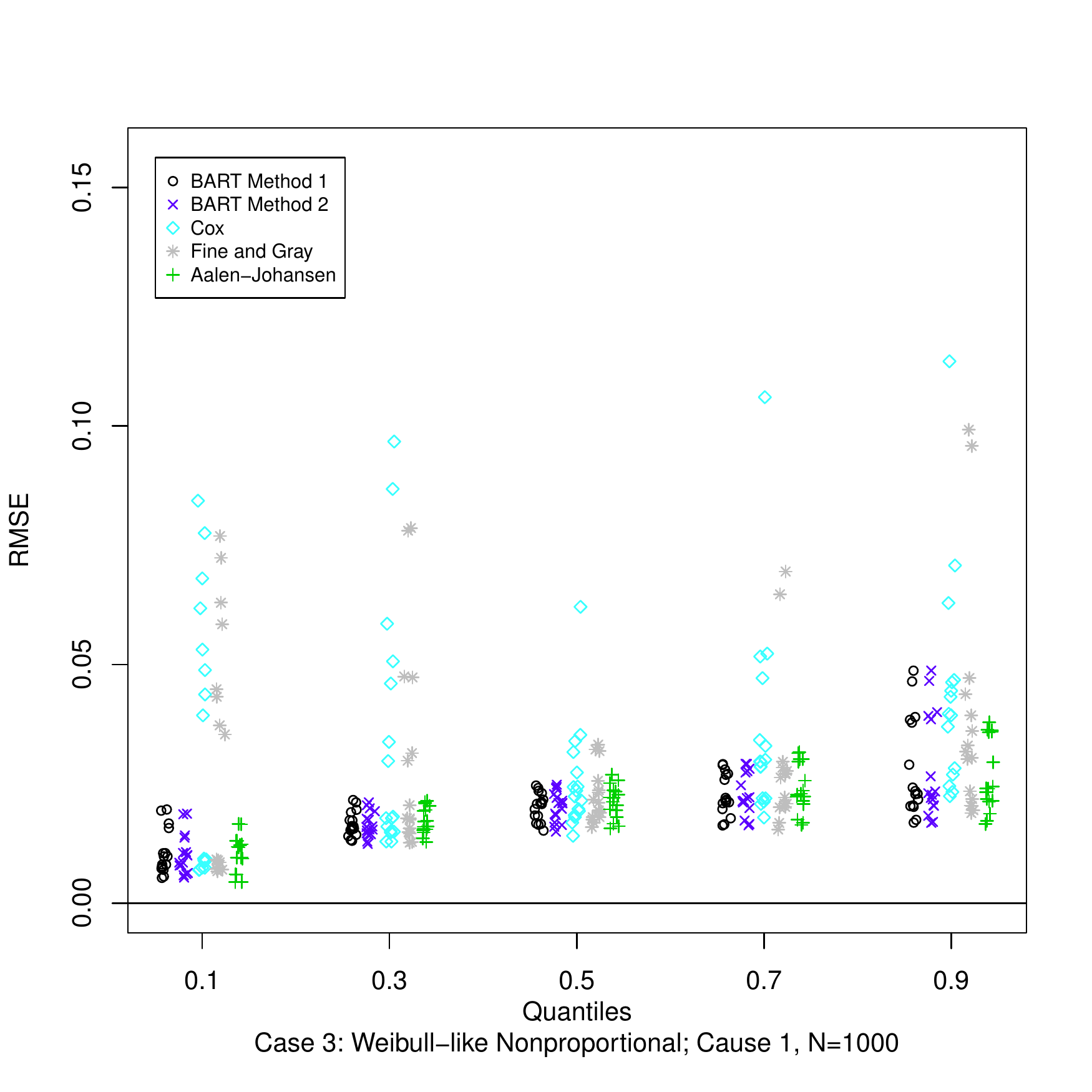}
\label{fig:3f} 
\end{minipage}% 
\caption{Bias (left) and RMSE (right) for case 3, $N=250$ (first row), $N=500$ (second row), and $N=1000$ (third row).  }\label{fig:3} \end{figure}

Results for bias and RMSE are shown for Cases~1, 2 and 3 in
Figures~\ref{fig:1}, \ref{fig:2} and \ref{fig:3} respectively.  In
terms of bias, for Case 1, as anticipated, the Cox model approach
generally has the smallest bias.  For Case 2, as anticipated, the Fine
and Gray method generally has the smallest bias.  For Case 3, BART
Method~2 generally has the smallest bias followed closely by BART
Method~1.  In terms of RMSE, for Case 1, generally all of the methods
are quite competitive with respect to RMSE.  Similarly for Case 2, all
of the methods are quite competitive with respect to RMSE.  For Case
3, the BART methods along with the Aalen-Johansen estimator, generally
have smaller RMSE than Cox and Fine and Gray.

\begin{figure}  
\begin{minipage}[b]{.5\linewidth} \centering\large 
\includegraphics[scale=0.35]{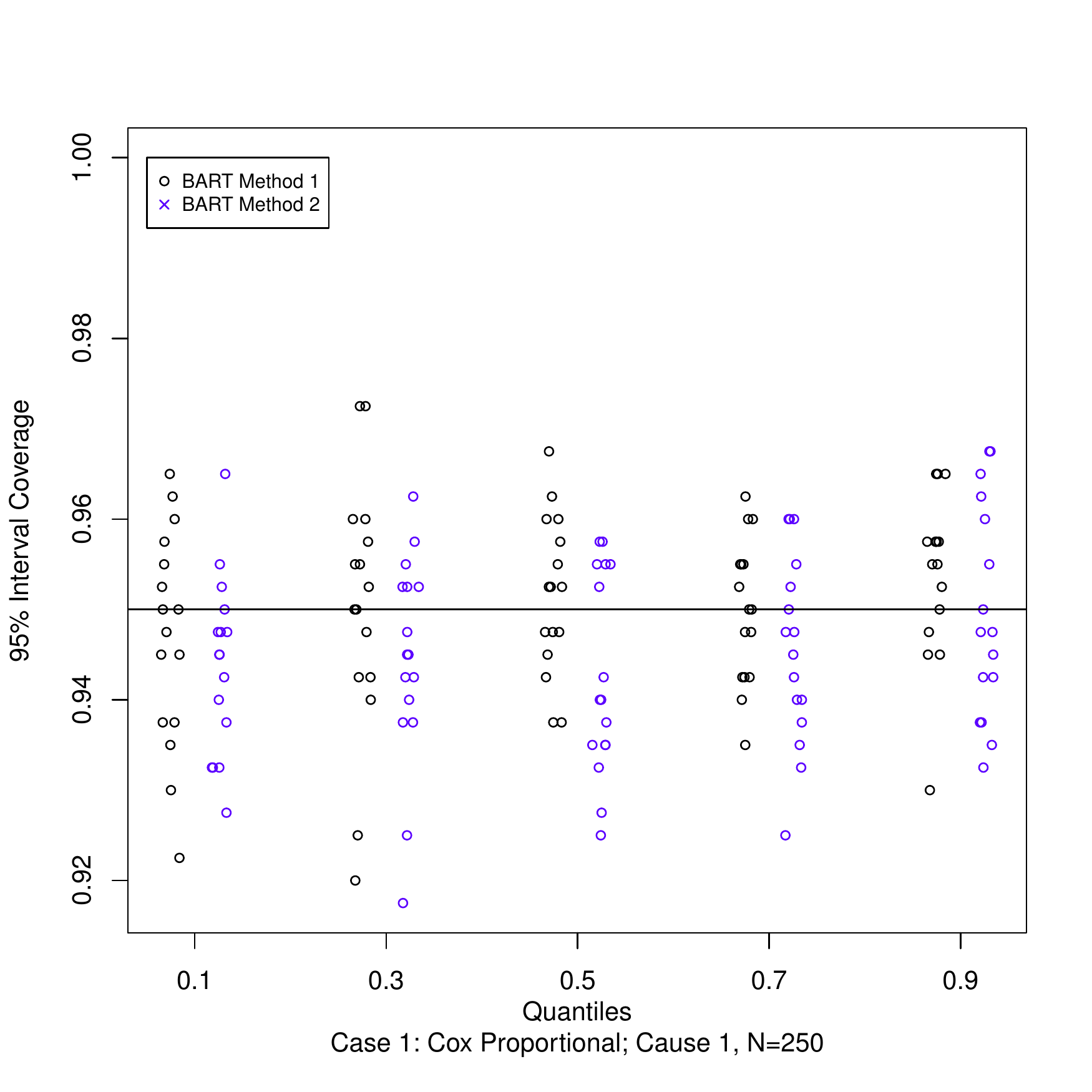}
 \label{fig:4a} \\
\includegraphics[scale=0.35]{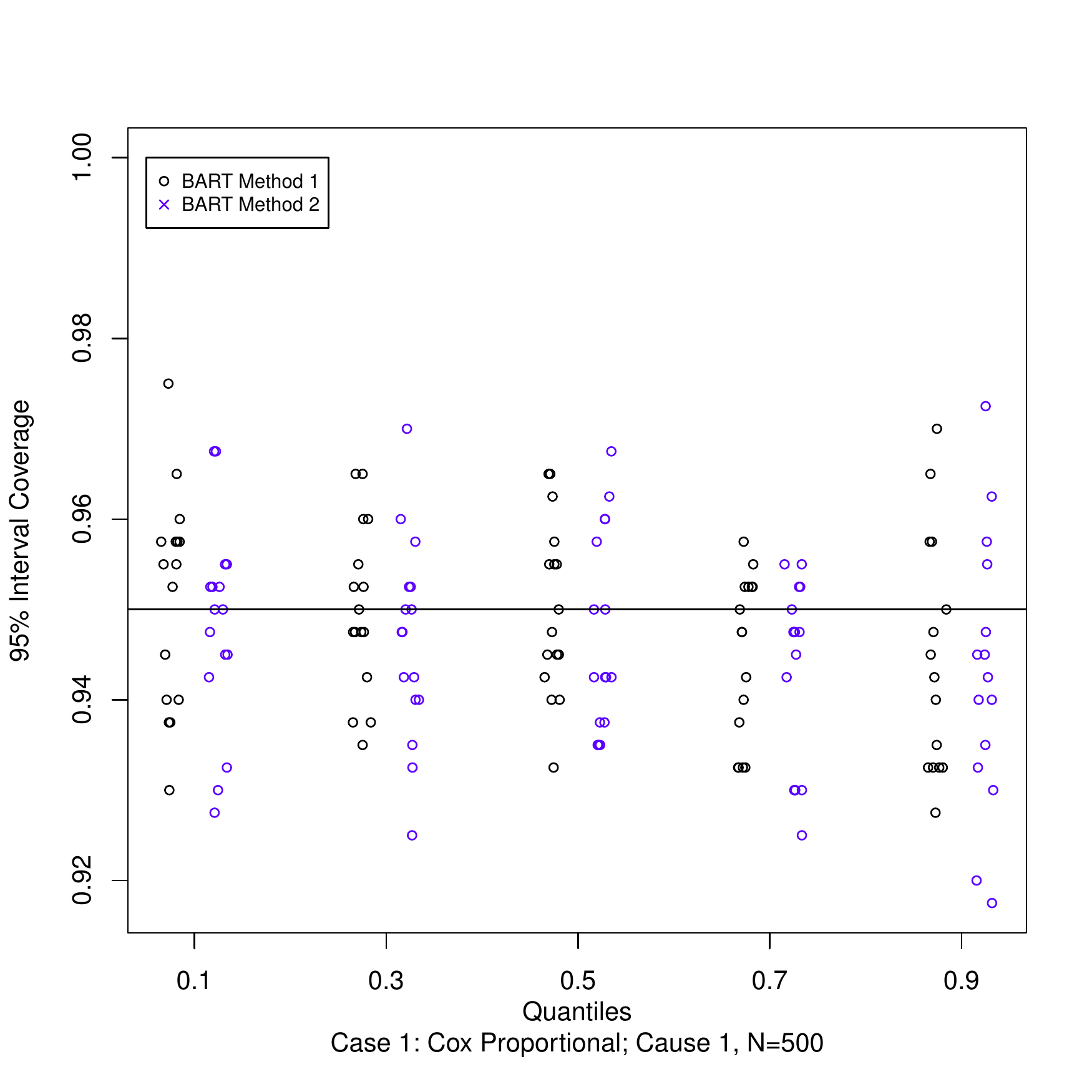}
 \label{fig:4c}  \\
\includegraphics[scale=0.35]{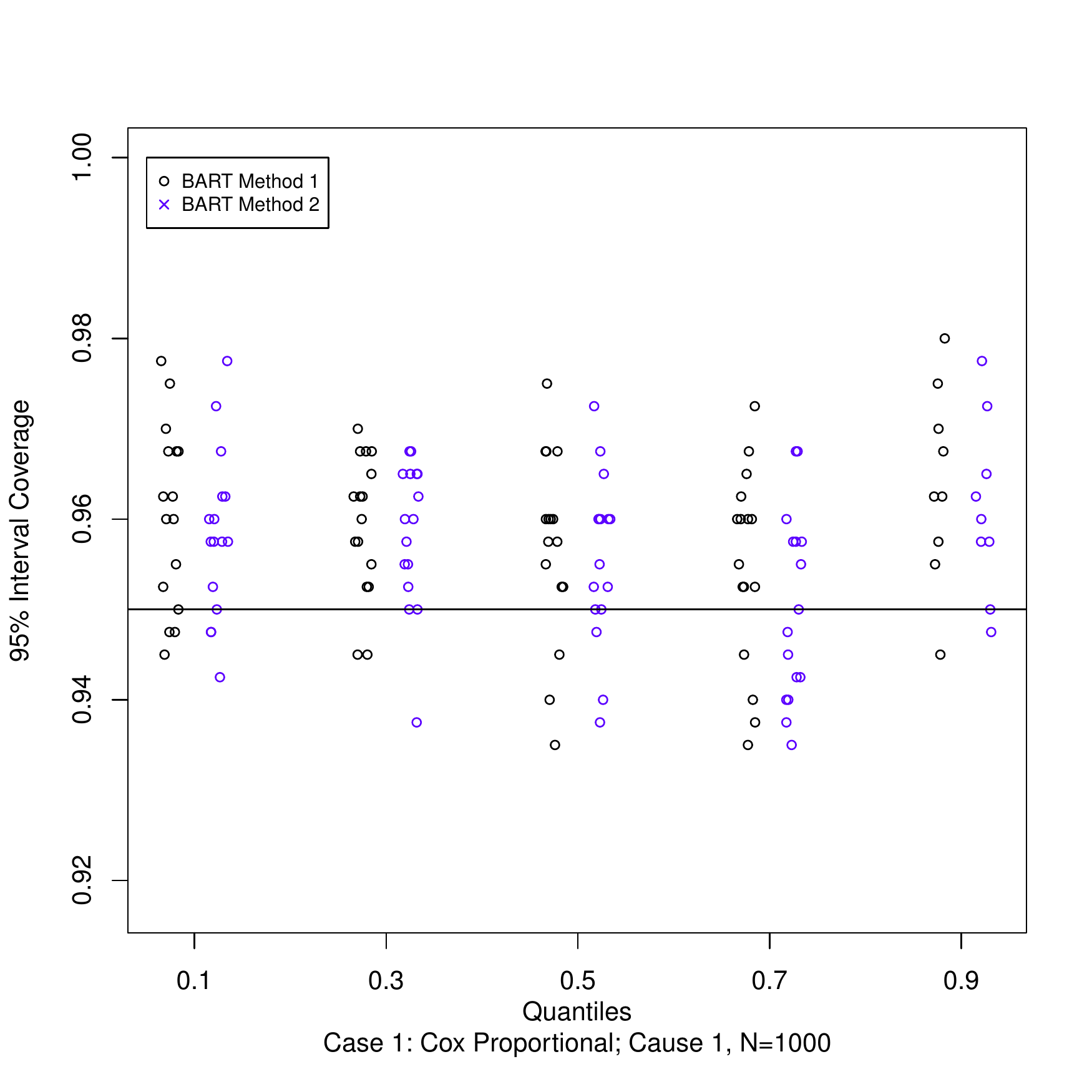}
 \label{fig:4e}
\end{minipage} 
\begin{minipage}[b]{.5\linewidth} \centering\large 
\includegraphics[scale=0.35]{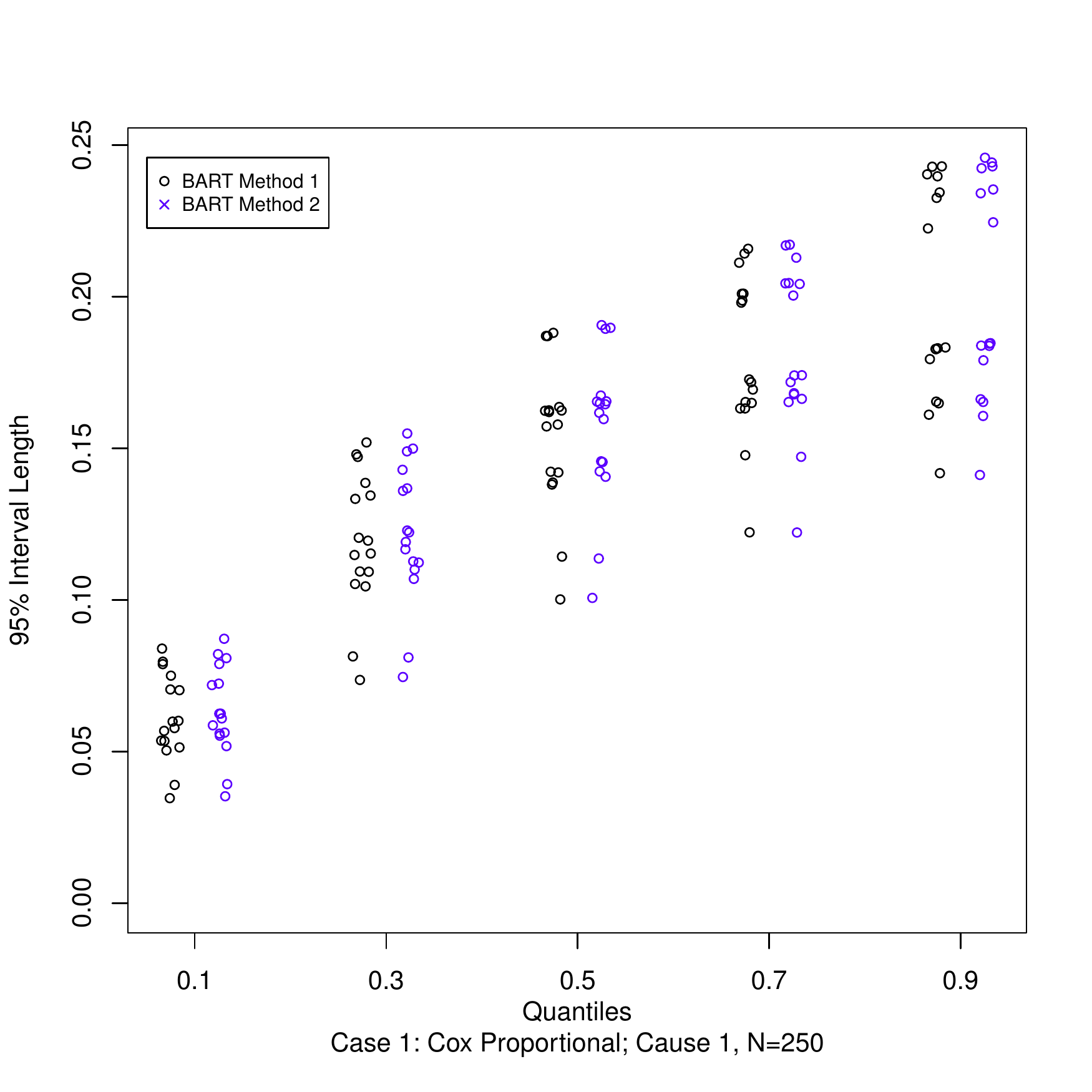} 
 \label{fig:4b} \\
\includegraphics[scale=0.35]{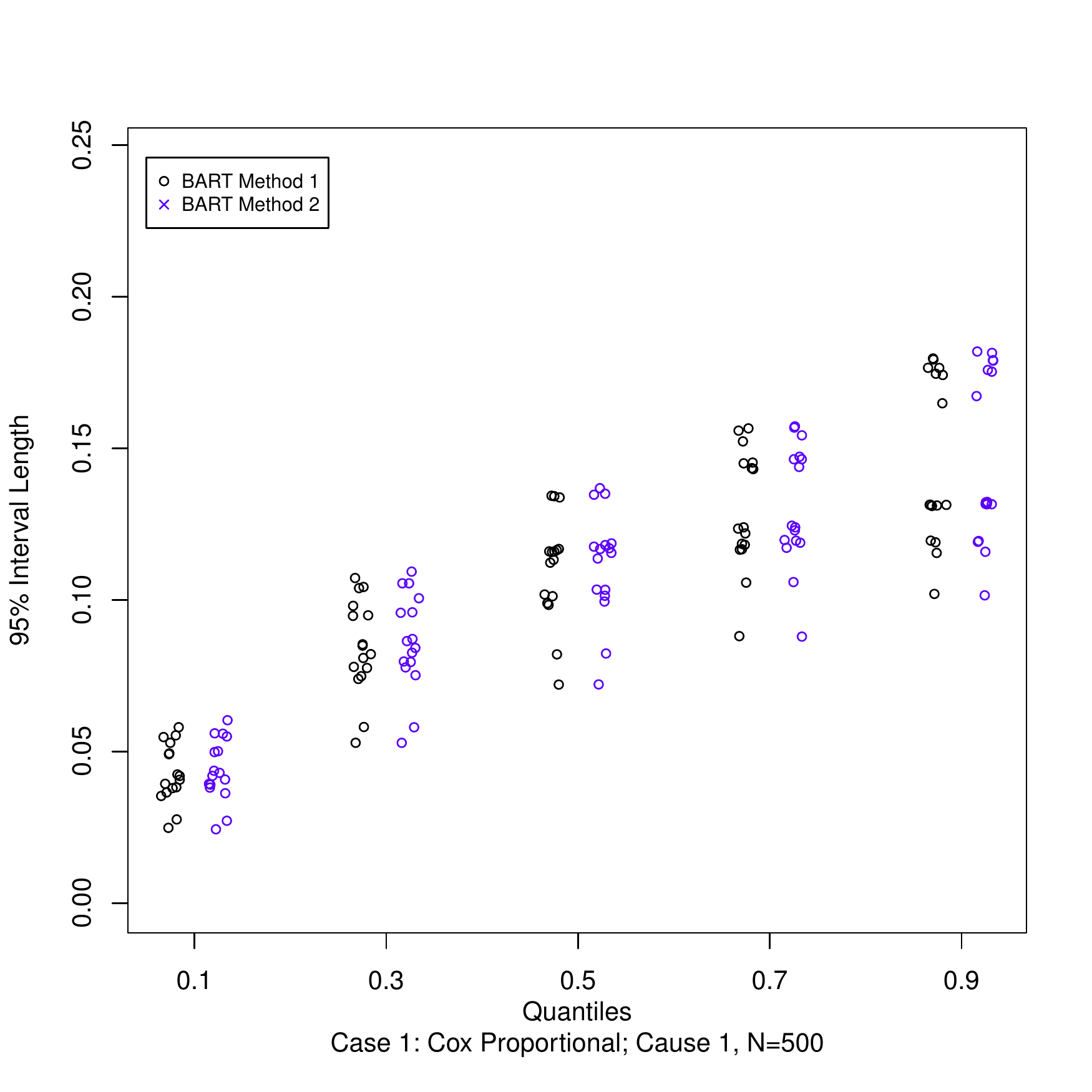}
\label{fig:4d} \\
\includegraphics[scale=0.35]{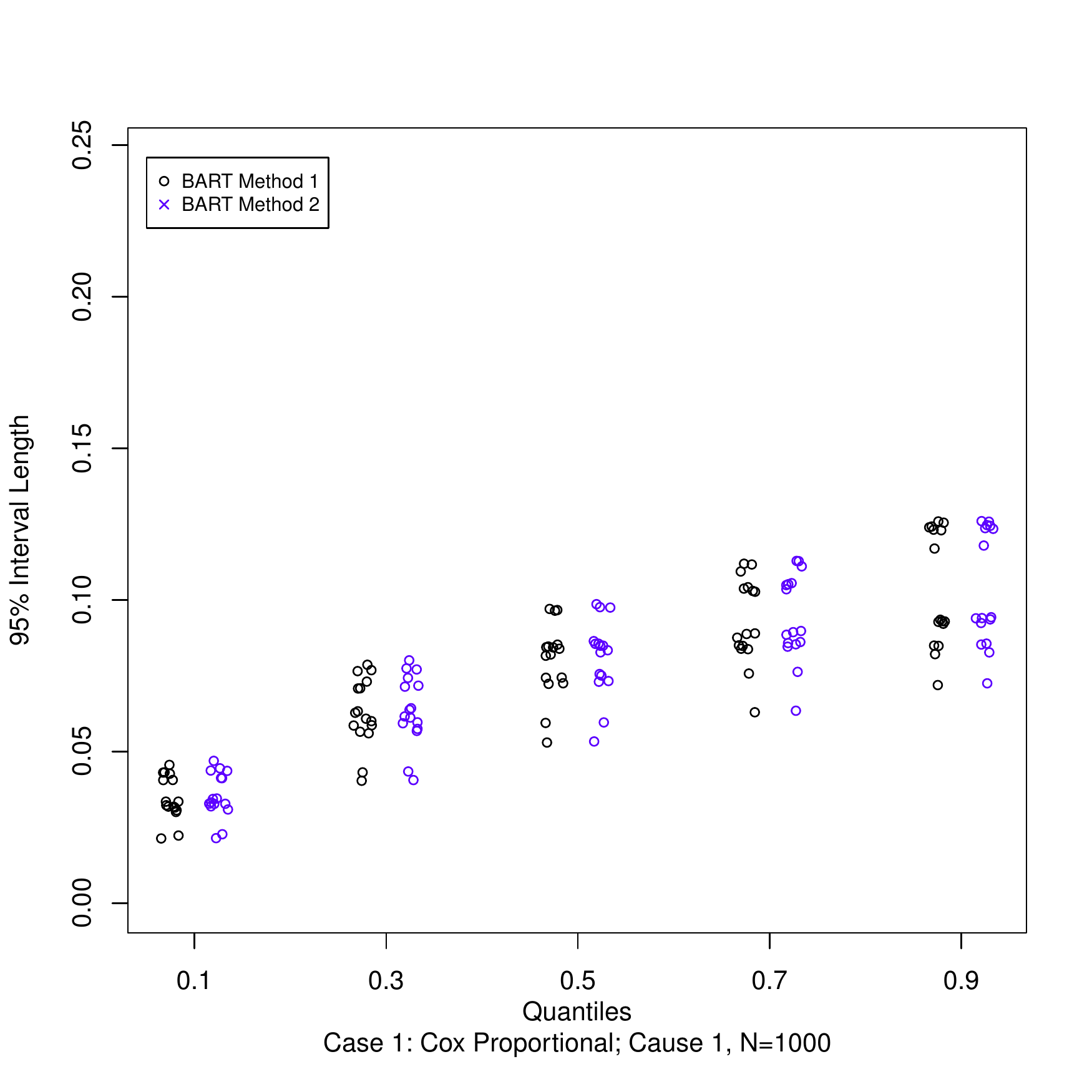}
\label{fig:4f} 
\end{minipage}% 
\caption{Coverage (left) and width (right) of 95\% posterior intervals for case 1, $N=250$ (first row), $N=500$ (second row), and $N=1000$ (third row).  }\label{fig:4} \end{figure}

\begin{figure}  
\begin{minipage}[b]{.5\linewidth} \centering\large 
\includegraphics[scale=0.35]{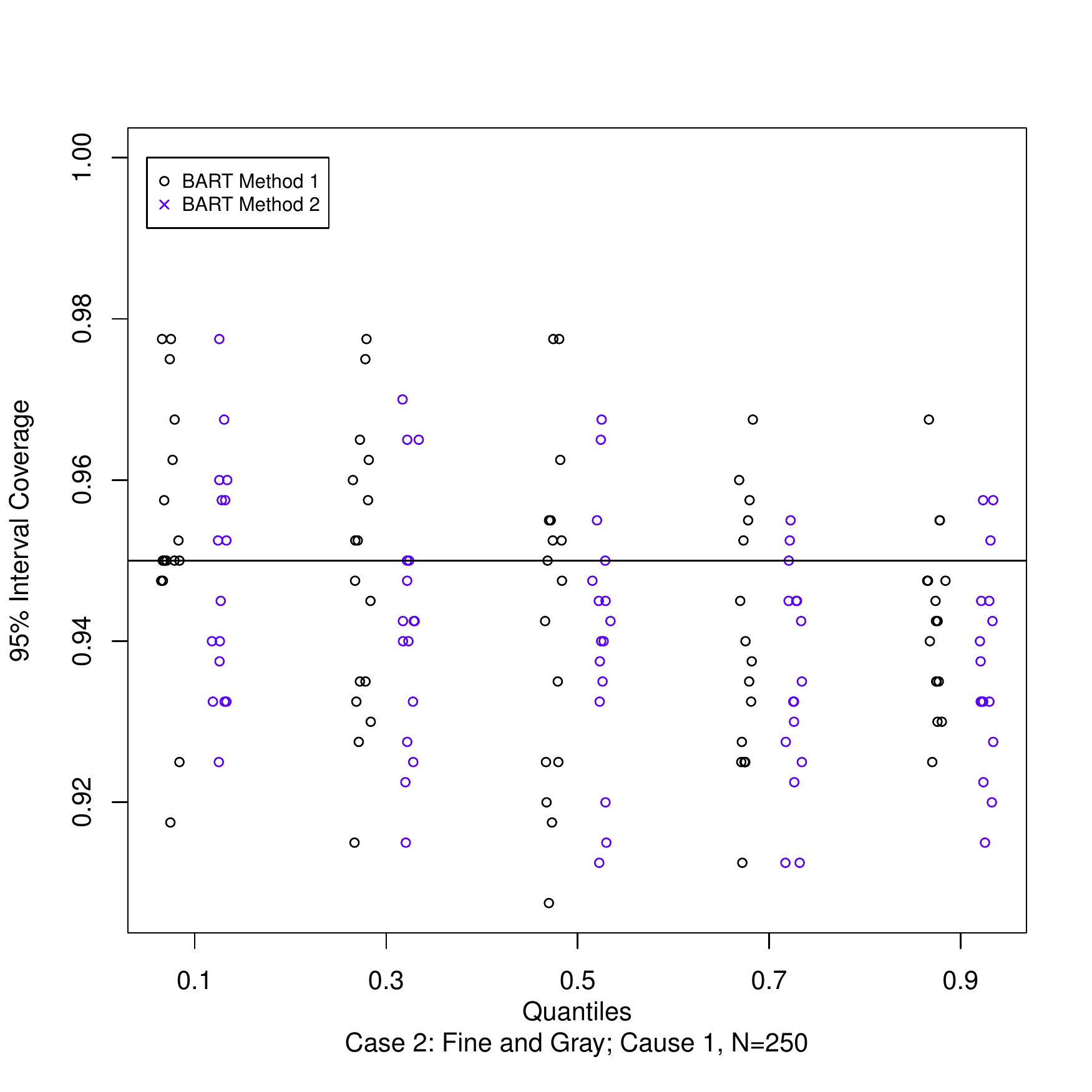}
 \label{fig:5a} \\
\includegraphics[scale=0.35]{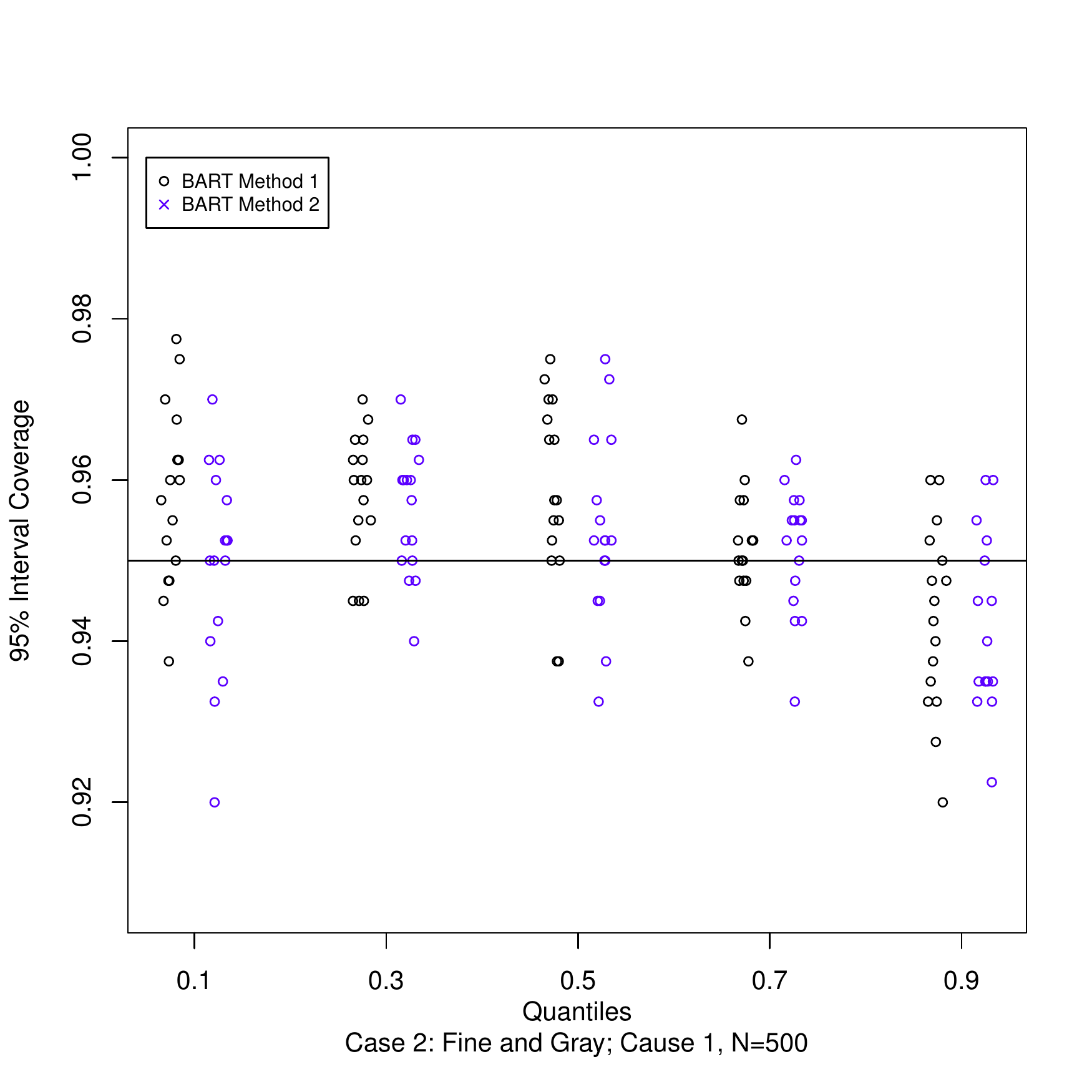}
 \label{fig:5c}  \\
\includegraphics[scale=0.35]{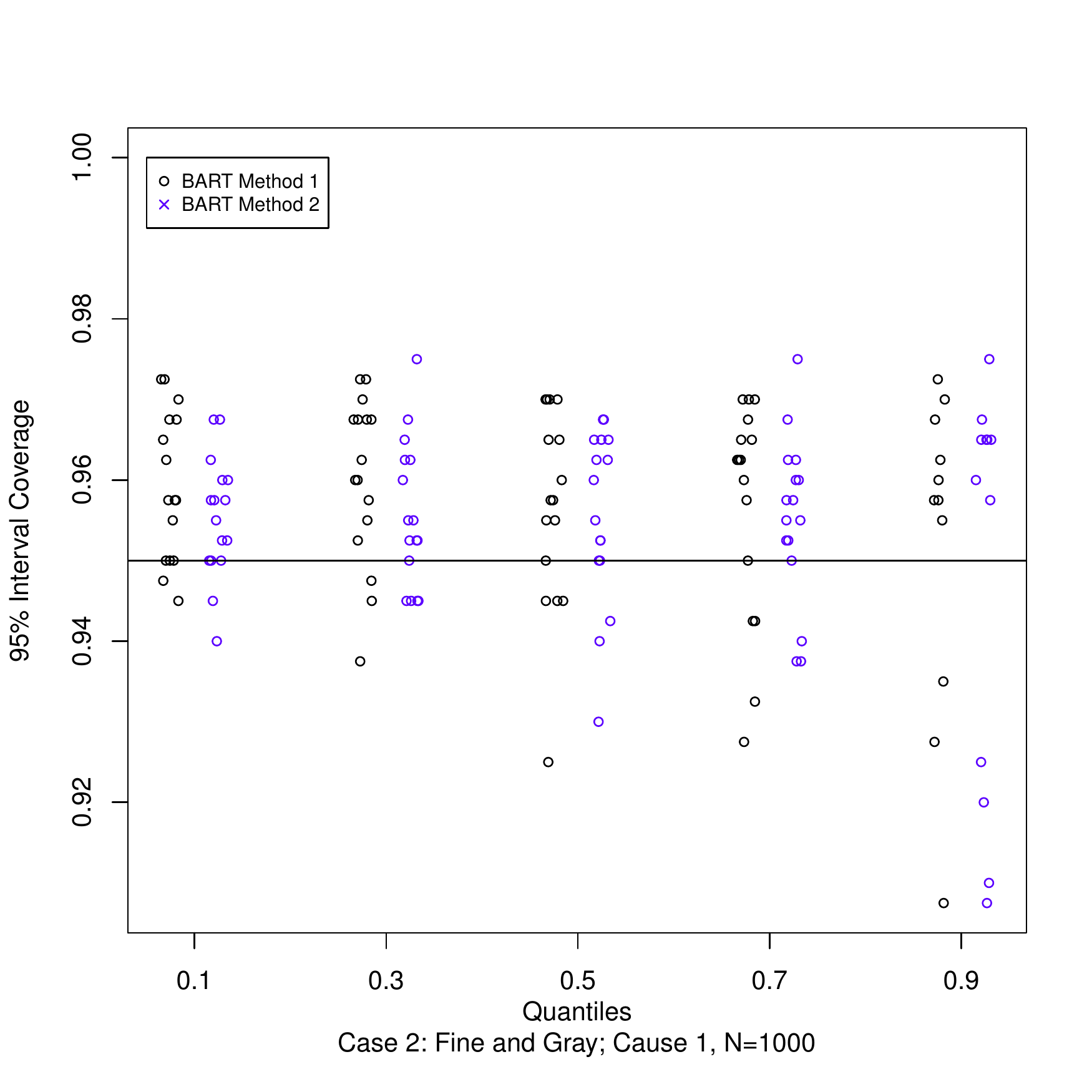}
 \label{fig:5e}
\end{minipage} 
\begin{minipage}[b]{.5\linewidth} \centering\large 
\includegraphics[scale=0.35]{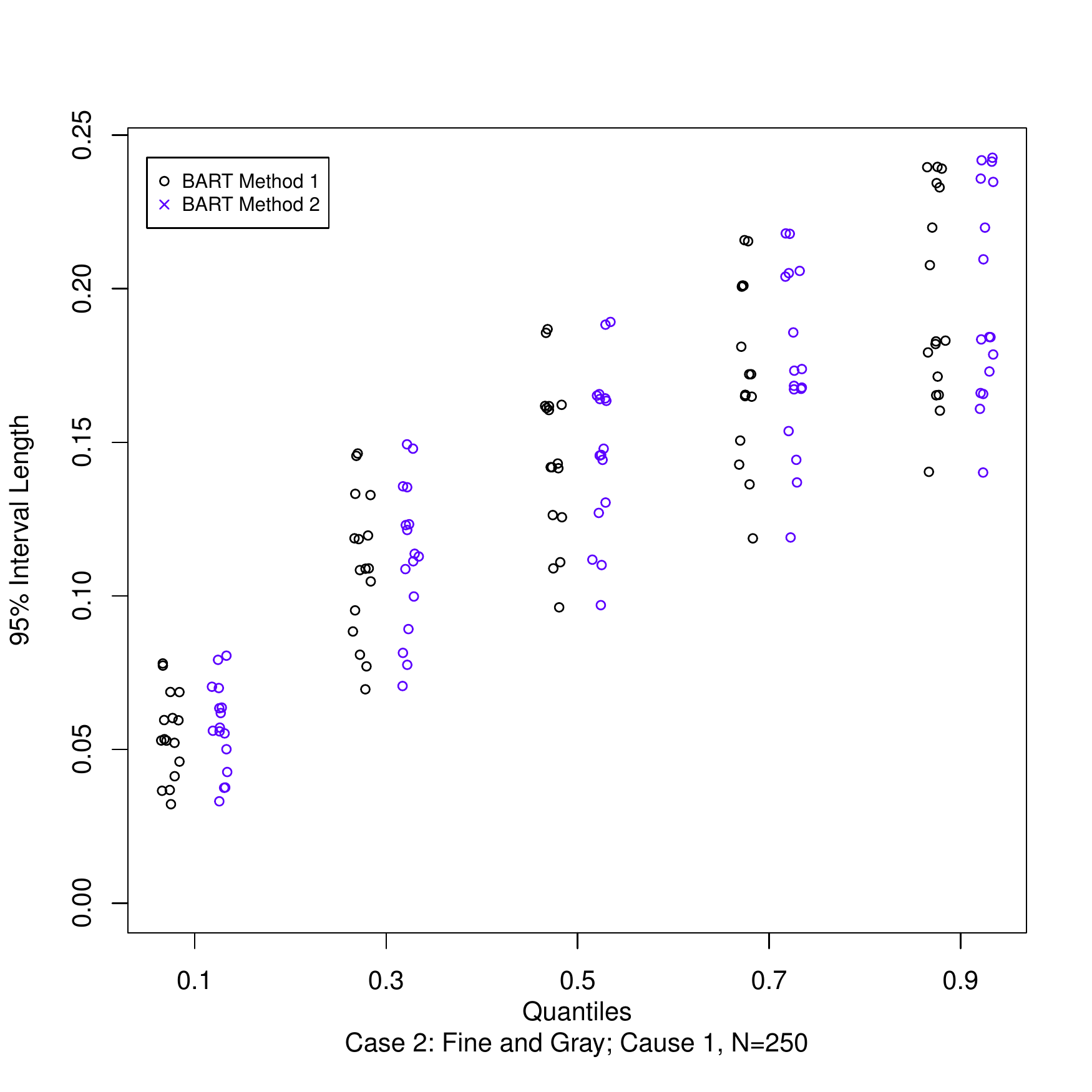} 
 \label{fig:5b} \\
\includegraphics[scale=0.35]{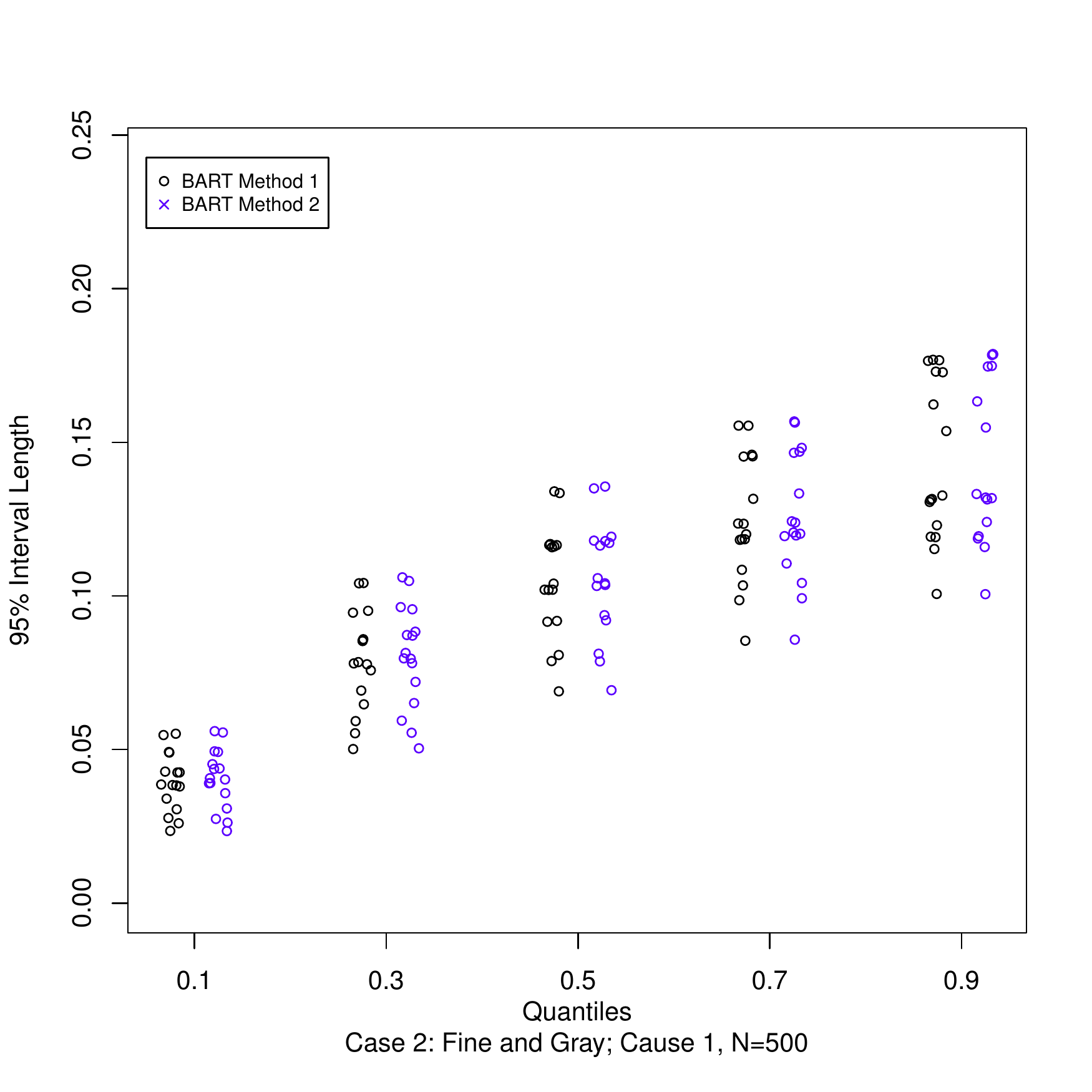}
\label{fig:5d} \\
\includegraphics[scale=0.35]{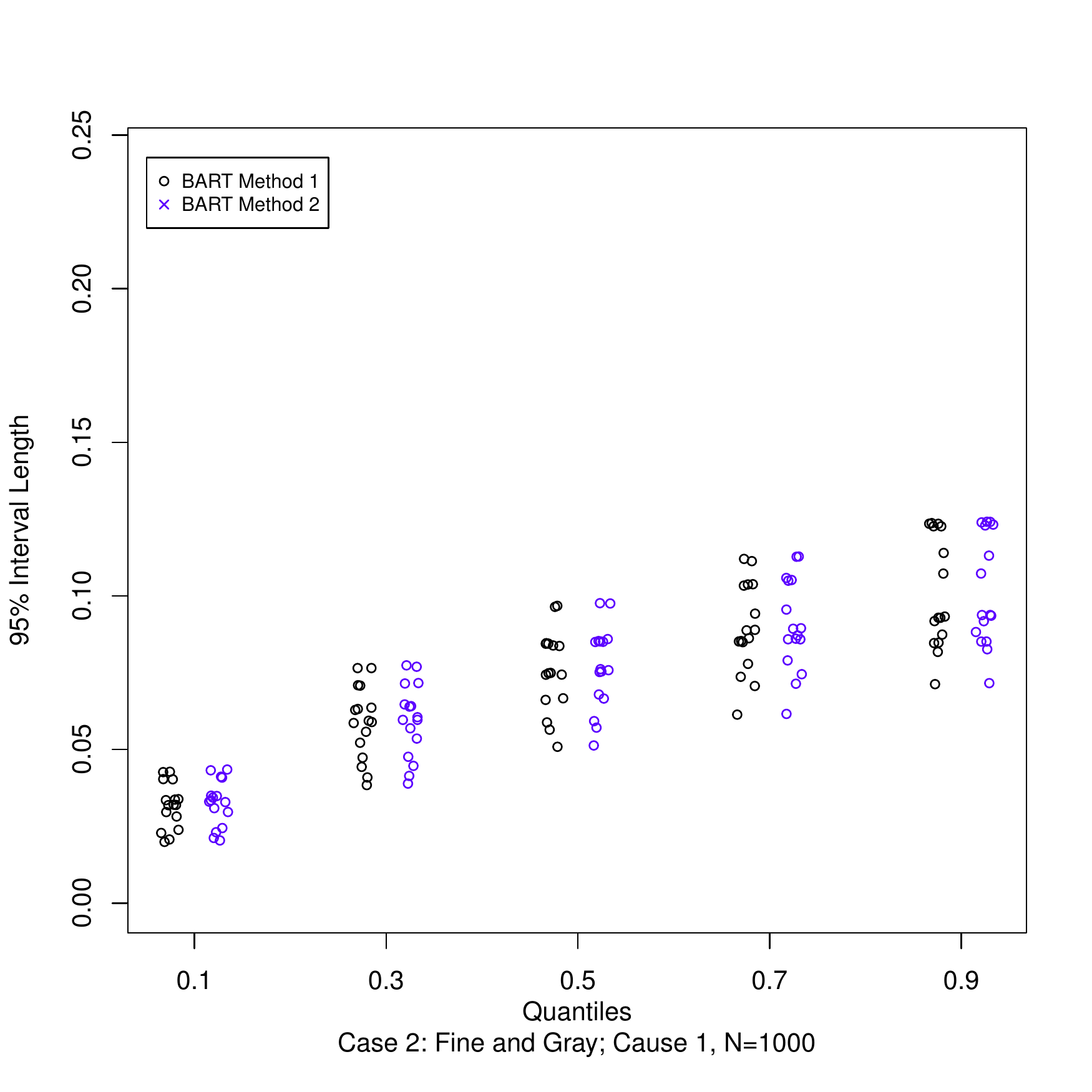}
\label{fig:5f} 
\end{minipage}% 
\caption{Coverage (left) and width (right) of 95\% posterior intervals for case 2, $N=250$ (first row), $N=500$ (second row), and $N=1000$ (third row).  }\label{fig:5} \end{figure}

\begin{figure}  
\begin{minipage}[b]{.5\linewidth} \centering\large 
\includegraphics[scale=0.35]{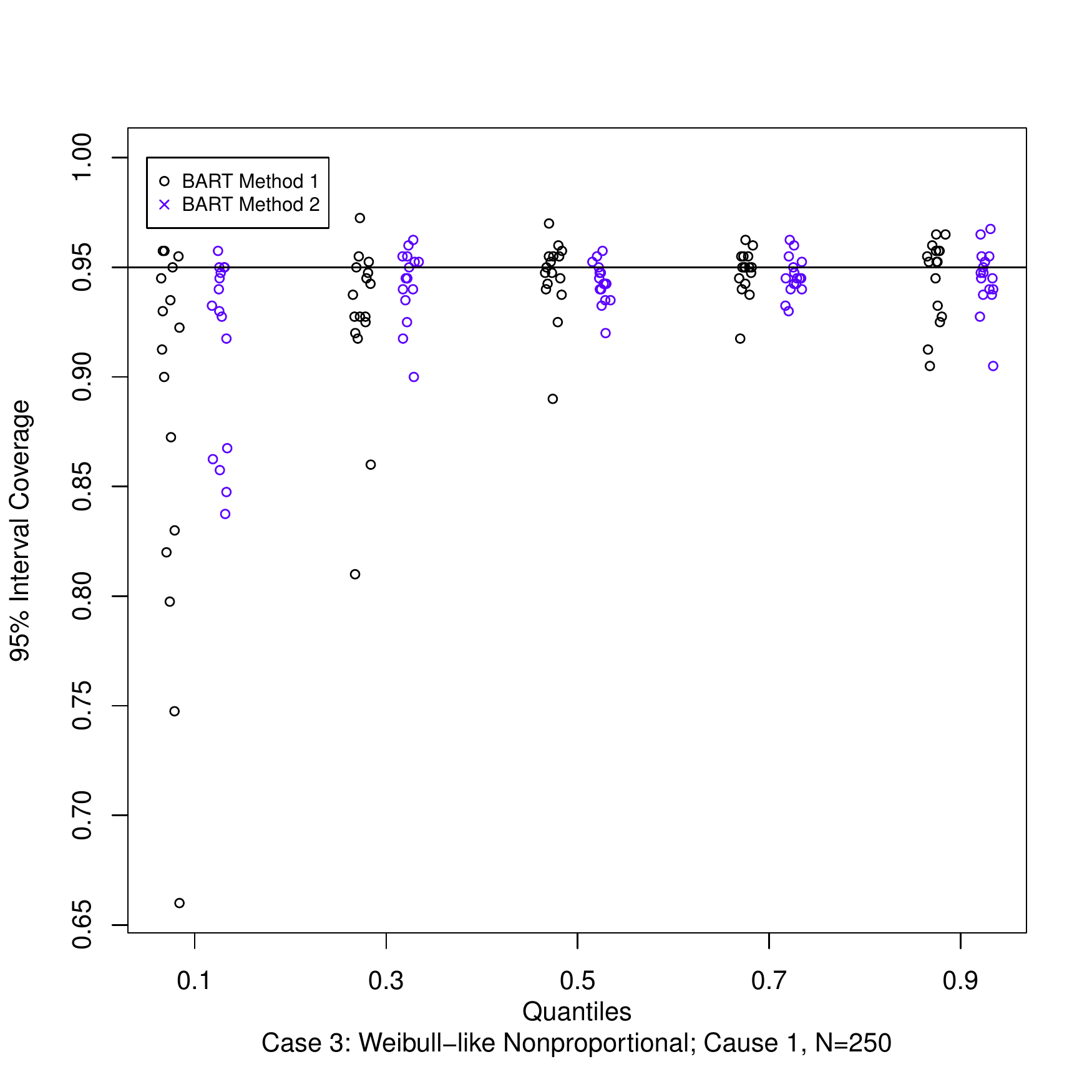}
 \label{fig:6a} \\
\includegraphics[scale=0.35]{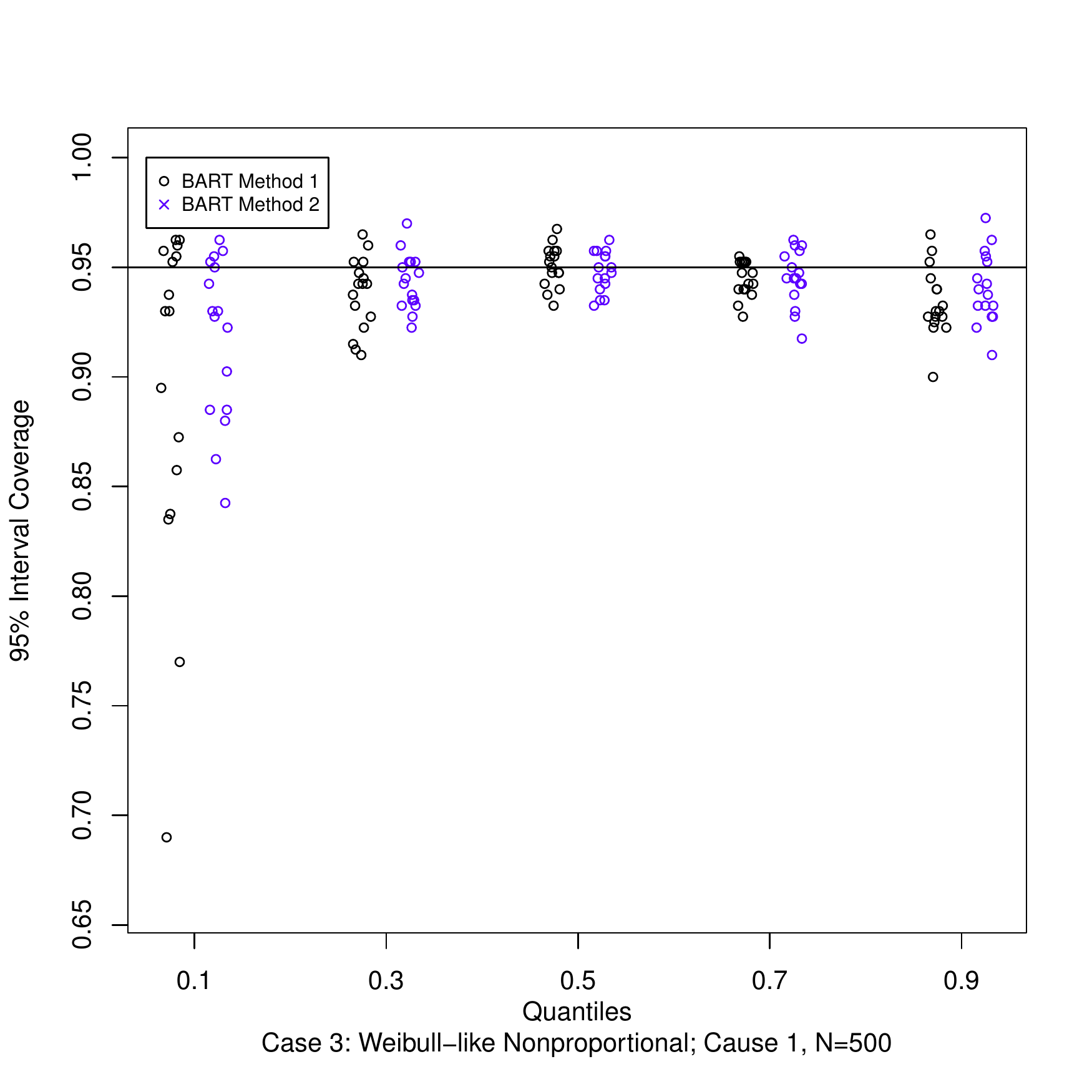}
 \label{fig:6c}  \\
\includegraphics[scale=0.35]{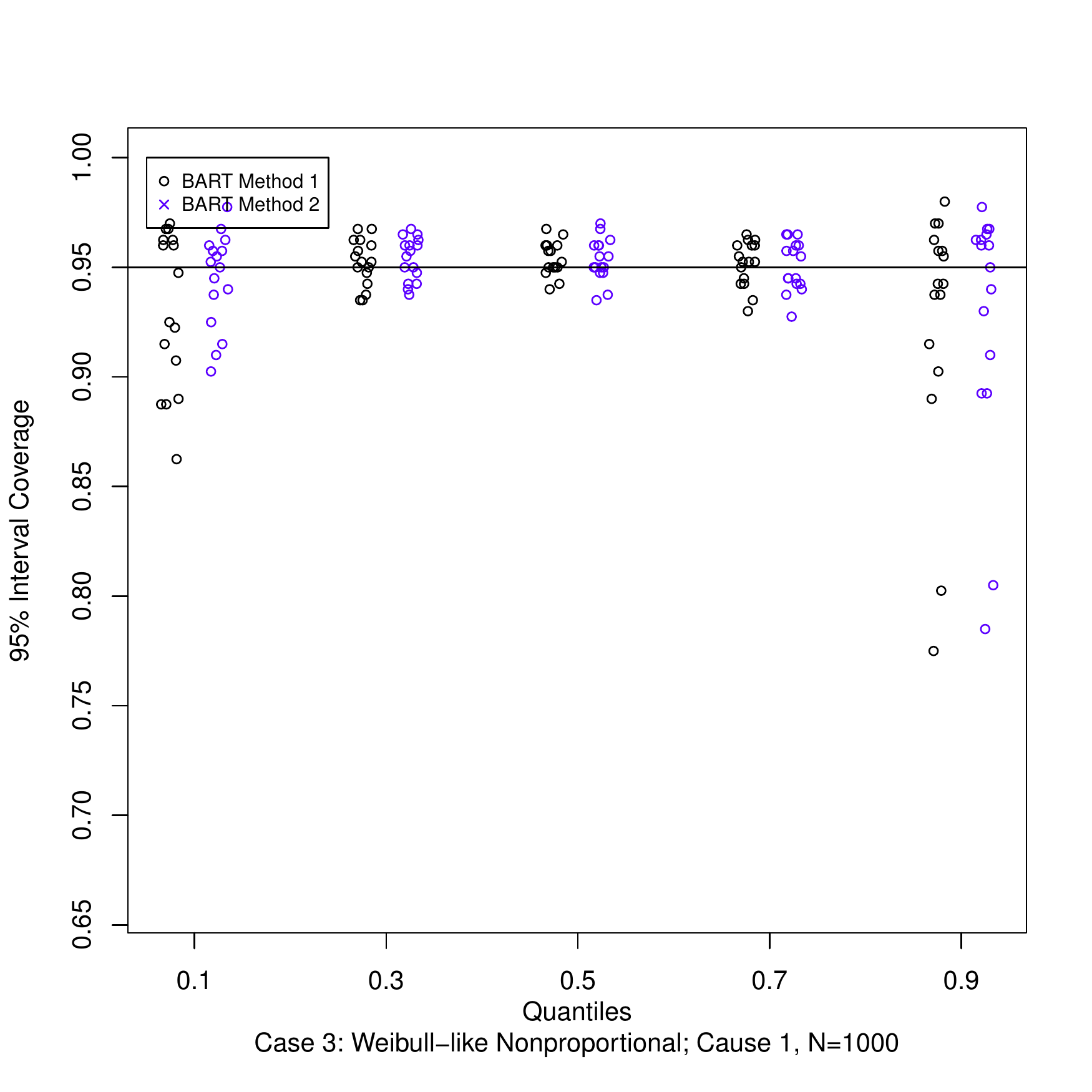}
 \label{fig:6e}
\end{minipage} 
\begin{minipage}[b]{.5\linewidth} \centering\large 
\includegraphics[scale=0.35]{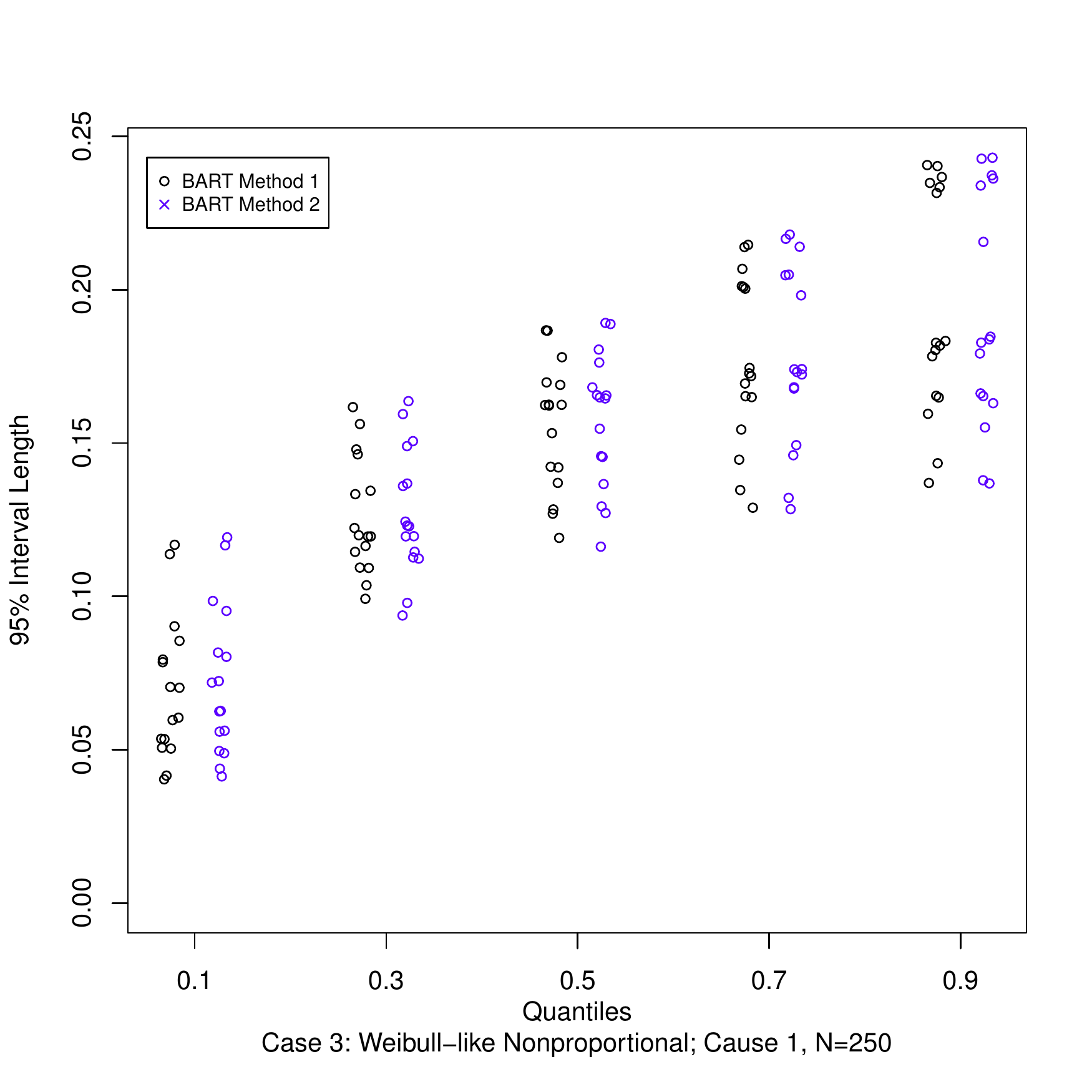} 
 \label{fig:6b} \\
\includegraphics[scale=0.35]{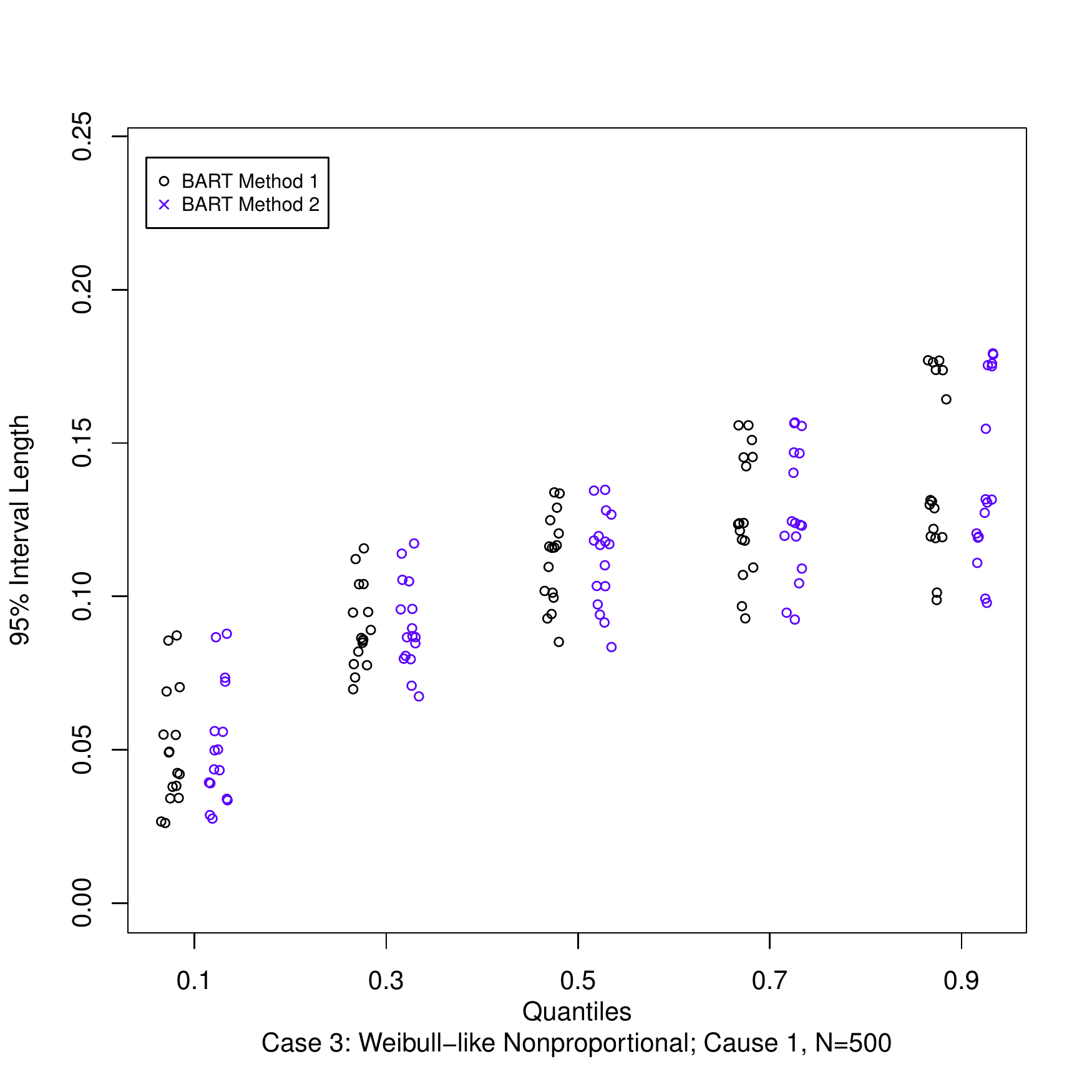}
\label{fig:6d} \\
\includegraphics[scale=0.35]{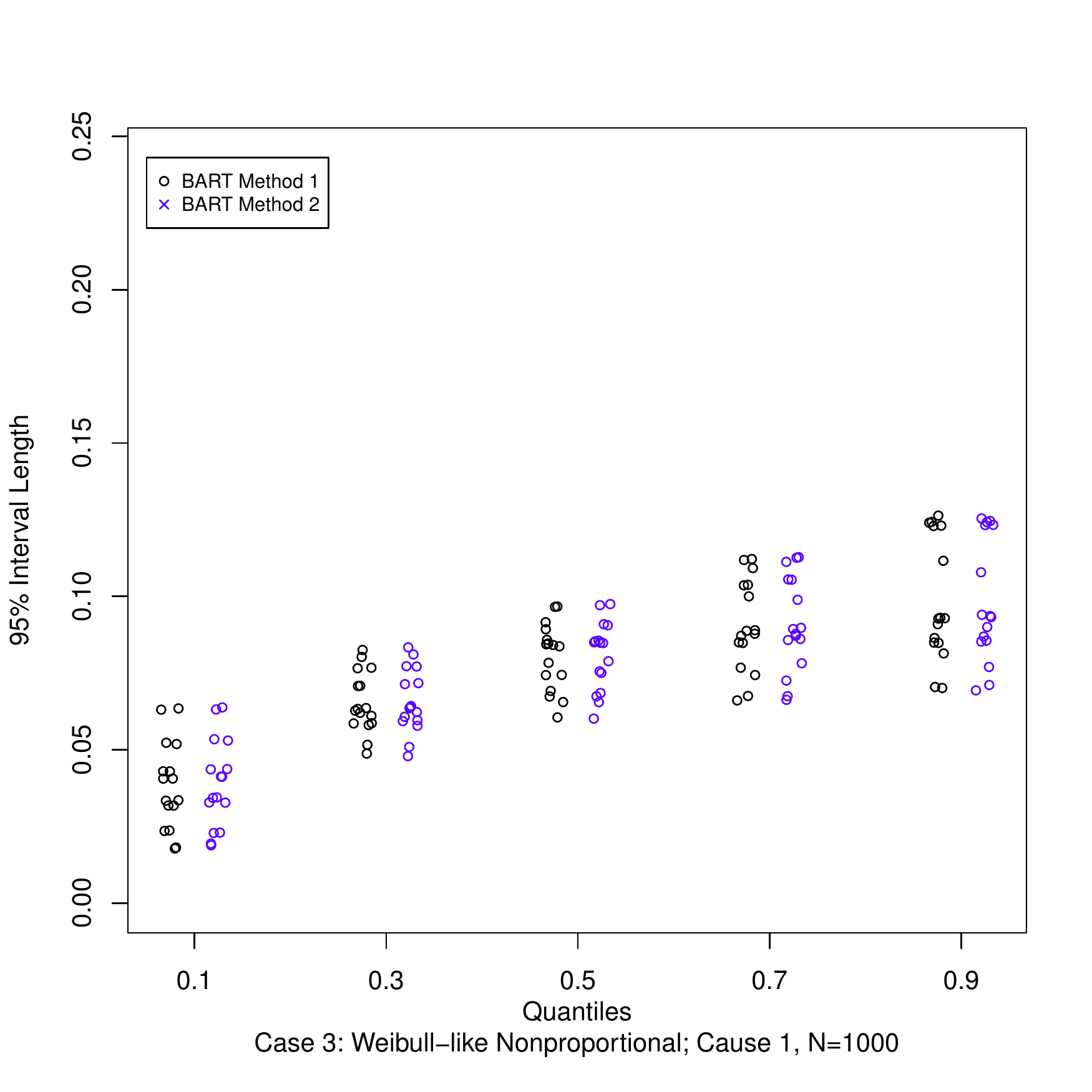}
\label{fig:6f} 
\end{minipage}% 
\caption{Coverage (left) and width (right) of 95\% posterior intervals  for case 3, $N=250$ (first row), $N=500$ (second row), and $N=1000$ (third row).  }\label{fig:6} \end{figure}

Results for coverage probabilities and interval length of 95\%
posterior intervals are shown for Cases~1, 2 and 3 in
Figures~\ref{fig:4}, \ref{fig:5} and \ref{fig:6} respectively.  For
all cases, both of the BART methods have good coverage.  There appears
to be little difference in the width of the intervals between the two
BART competing risk approaches.  In summary, the BART methods perform
comparable to the best method for each case considered in the two
sample setting.  This establishes the validity of the BART competing
risks methodology as a flexible nonparametric estimator of the
cumulative incidence function even in the presence of a binary
covariate.  No noticeable differences in performance were seen 
between method 1 and method 2; however, method 1 has an advantage in
terms of computation time because the second constructed data set used
for the second BART function is substantially smaller (as can be
clearly seen from $u_i$ in Table~\ref{dataconstruct}).

\subsection{Complex regression setting}

While the above simulation establishes BART as a nonparametric
estimator of the cumulative incidence function in the presence of a
binary predictor, in practice, we are more interested in utilizing
these approaches for modeling of competing risks data with more
complex regression relationships.  In this section, we demonstrate the
performance of the proposed methods in a complex regression setting,
and benchmark it against Random Survival Forests
\citep{IshwGerd14,IshwKoga18}.  We generated two simulated data sets
for each of the sample sizes; $N=500, 2000, 5000$; for one data set we
generated a small number of covariates, $P=10$, and the other we
generated a large number of covariates, $P=1000$.  We base this
setting on the Fine and Gray model \citep{FineGray99} since it
provides a direct analytic expression for the cumulative incidence
functions, and we only show the results of cause~1 for brevity.
Because we are examining the impact of high dimensional predictors, we
compare two variants of BART Method~1 against Random Survival Forests
(RSF).  The first variant is standard BART which chooses among the
variables with a uniform prior.  The second variant, which we call
DART, substitutes a sparse Dirichlet prior for variable selection.

The basics of this setting are provided in Case 2 above, except that
in the cumulative incidence expression \ref{fgcif}, we set $p_0=0.2$
and replace $x\beta_1$ with $f(\bm{x})$ (which was inspired by
Friedman's five-dimensional test function \citep{Frie91}):
$f(\bm{x})= 0.5 \sin(\pi x_{1}x_{(0.5P+1)})+
x_2^2 + 0.5 x_{(0.5P+2)}+0.25 x_{3}^2-1.25$ where\\
$x_{j} ~ \U{-1, 1}\ j=1, \., 0.5P$ and
$x_{j'} ~ \U{\{-1, 1\}}\ j'=0.5P+1, \., P$.\\
Note that this prescription provides $f(\bm{x}) \in [-1, 1]$.

The models are fit to the randomly generated training data and applied
to an independent test sample of size 500 in order to plot the
predicted cumulative incidence against the true CIF at select time
points based on quantiles of the observed cause 1 event times.  Lin's
concordance coefficient (labeled $R^2$)\citep{LinHeda02} was also
provided to summarize the agreement between the predicted and true
cumulative incidence function for cause 1.

\begin{figure}
\includegraphics{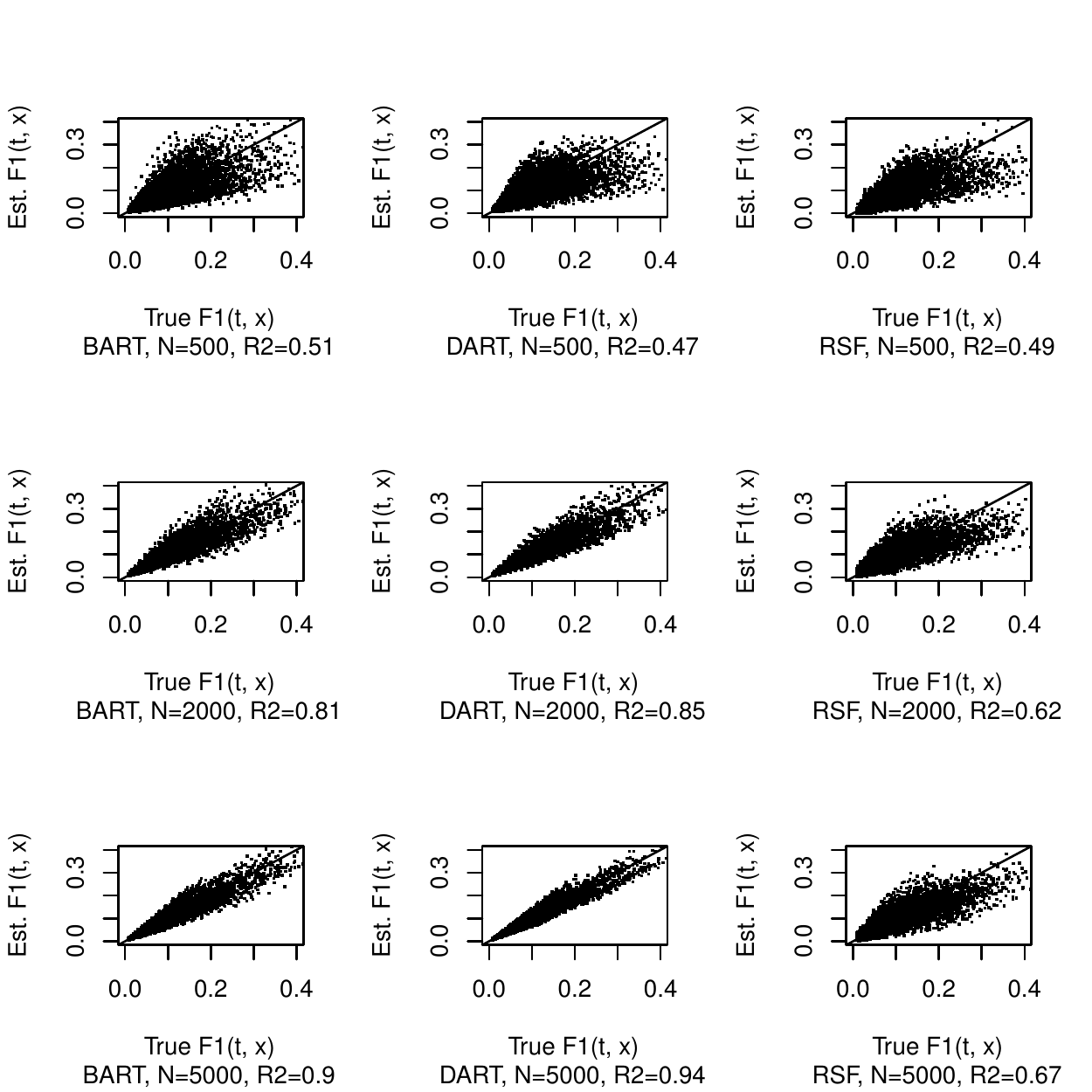}
\caption{Predicted vs. true $F_1(t,x)$ for BART, DART, and RSF, with $P=10$, at sample sizes of $N=500, 2000, 5000$.  At $N=500$, all three methods have roughly
  equivalent $R^2$ around 0.5.  At $N=2000,5000$, DART has a
  slight advantage over BART and DART/BART have better performance
  than RSF.  \label{case6-10}}
\end{figure}

\begin{figure}
\includegraphics{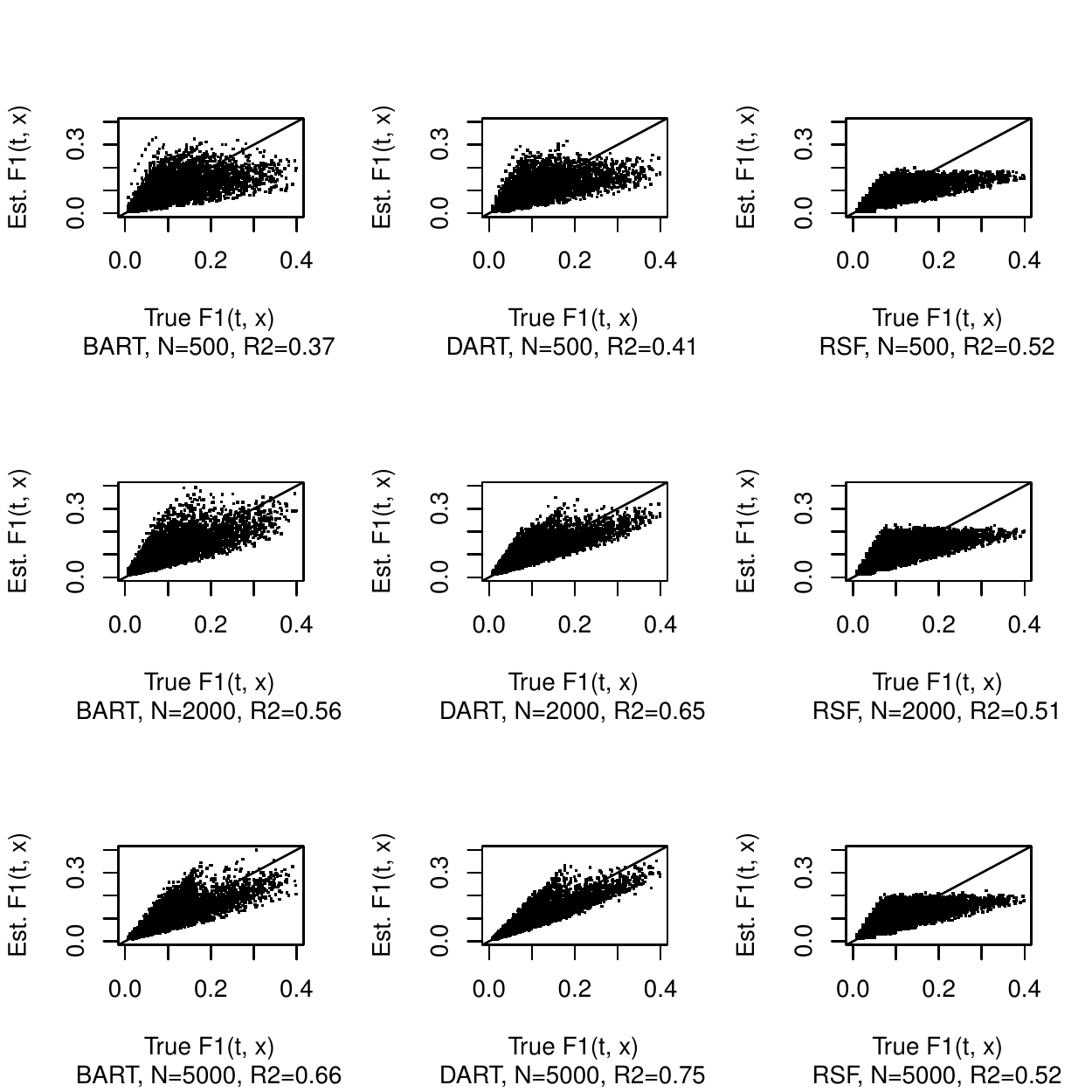}
\caption{Predicted vs. true $F_1(t,x)$ for BART, DART, and RSF, with $P=1000$, at sample sizes of $N=500,2000,5000$.  At $N=500$, the RSF method has an advantage.
  At $N=2000,5000$, DART has an advantage over BART
  and DART/BART have better performance than RSF.  \label{case6-1000}}
\end{figure}

The results for $P=10$ and $P=1000$ are shown in
Figures~\ref{case6-10} and \ref{case6-1000} respectively.  For $P=10$,
at $N=500$, all three methods have roughly equivalent $R^2$ around
0.5.  When we get to $N=2000$, DART has a slight advantage over BART
and DART/BART have better performance than RSF.  Similar results were
obtained at $N=5000$.

For $P=1000$, at $N=500$, the RSF method has an advantage.  However,
when we get to $N=2000$, DART has an advantage over BART and DART/BART
have better performance than RSF.  Similar results were obtained at
$N=5000$.  Surprisingly, RSF's concordance is consistently 0.5
regardless of sample size, and all covariate combinations seem to
converge on the same limiting cumulative incidence.  This may be due
to the inability of RSF to adapt to sparsity as reported in
\citep{Line17} without an explicit strategy for variable selection.
Since only $\sqrt{P}$ variables are checked at each split, the
likelihood of finding and splitting on important variables is low,
leading to mostly random splits which would have a similar limiting
cumulative incidence.  We speculate that this could be mitigated by
incorporating variable selection strategies based on variable
importance measures directly into the algorithm.

\section{Application: hematopoietic stem cell transplantation data}
\label{Real-ex}

In this section, we apply the proposed BART competing risks method to
a retrospective cohort study data set looking at the outcome of
chronic graft-versus-host disease (cGVHD) after a reduced intensity
hematopoietic cell transplant (HCT) from an unrelated donor
\citep{EapeLoga15} between the years 2000 to 2007.  Development of
cGVHD is the event of interest while death prior to development of
cGVHD is the competing event.  Patients with missing covariate data
were removed to facilitate demonstration of the methods, so the
results should be considered as an illustration of the methods rather
than a clinical finding.  A total of 427 cGVHD events and 324
competing risk events occurred in the 845 patients in the cohort.
Thirteen covariates were considered in the analysis, including age,
matched ABO blood type, year of transplant, disease/stage, matched
human leukocyte antigens (HLA), graft type, Karnofsky Performance
Score (KPS), cytomegalovirus (CMV) status of the recipient,
conditioning regimen, use of in vivo T-cell depletion,
graft-versus-host disease (GVHD) prophylaxis, matched donor-recipient
sex and donor age, resulting in a total of 21 predictors in the X
matrix. More details on the variables are available in
\citep{EapeLoga15}.  The time scale was coarsened to weeks rather than
days to reduce the computational burden.

\begin{figure}

\centering
\includegraphics[scale=0.45]{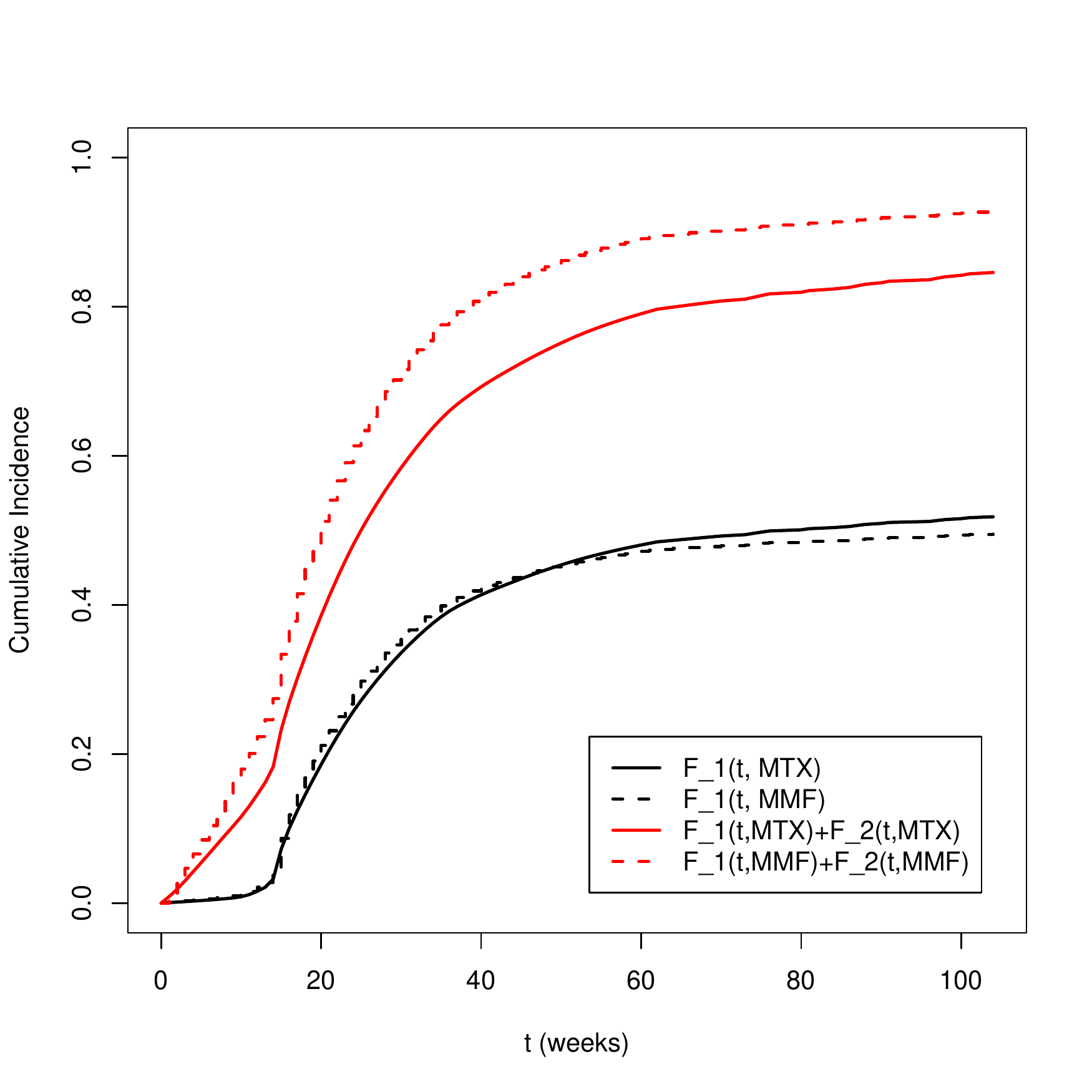}
\caption{Partial Dependence Stacked Cumulative Incidence Functions for two different GVHD prophylaxis strategies: MTX based or MMF based.  Bottom line is the CIF for cGVHD, while top line represents the sum of the CIF for cGVHD and for the competing risk of death.  
\label{Fig:PDCIF}}
\end{figure}

The BART competing risks Method~2 was fit to this data set with 200
trees, %the Dirichlet sparse prior 
and the default settings for the
rest of the prior settings, using a burn-in of 100 draws and thinning
by a factor of 10, resulting in 2000 draws from the posterior
distributions for the cumulative incidence function given covariates.
%(RS: WHY 200?  THE DEFAULT IS 50). 
Based on our simulation studies, we
expect Method 1 to yield similar results, so we do not show it here.
Partial dependence cumulative incidence functions can be obtained as
in equation \eqref{eq:pdcif} for a particular subset of covariates.
%{\bf NEED TO DEFINE PARTIAL DEPENDENCE CIF}.  
These can be interpreted
as a marginal or average cumulative incidence function for that
covariate level, averaged across the observed distribution of the
remaining covariates.  In the left panel of Figure~\ref{Fig:PDCIF}, we
show the stacked partial dependence cumulative incidence functions for
each of two GVHD prophylaxis strategies, Methotrexate (MTX) based or
Mycophenolate Mofitil (MMF) based.  For each strategy, the CIF for
cGVHD are shown as the bottom line, while the sum of the CIF for cGVHD
and for death prior to cGVHD are shown as the upper line.  These
indicate that while there is very little difference in the incidence
of cGVHD between these strategies overall, there seems to be a higher
rate of death without cGVHD in the MMF group.

\begin{figure}
\centering
\includegraphics[angle=270,scale=0.4]{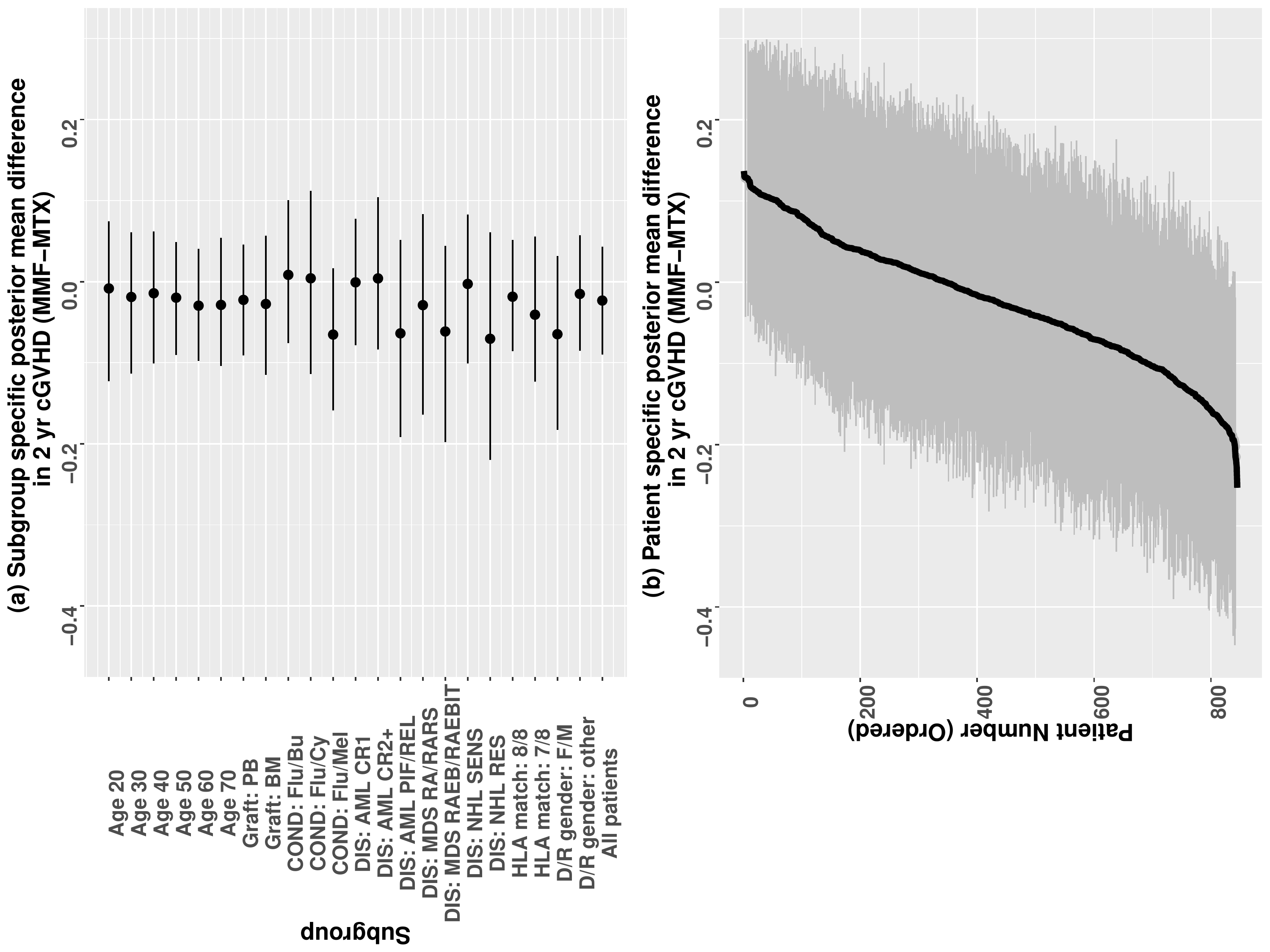}
\caption{Plots of the difference in 2 year CIF for cGVHD, with 95\%
  posterior interval, by (a) clinically defined subgroup, and (b)
  individual patient (ordered by mean difference).  Subgroups are by
  age, graft type (PB=Peripheral Blood, BM=Bone Marrow), conditioning
  regimen (Flu=Fludarabine, Bu=Busulfan, Cy=Cyclophosphamide,
  Mel=Melphalan), disease/stage (AML=Acute Myelogenous Leukemia,
  MDS=Myelodysplastic Syndrome, NHL=Non-Hodgkin's Lymphoma,
  CR=Complete Remission, PIF=Primary Induction Failure, REL=Relapse,
  RA=Refractory Anemia, RAR=RA with Ringed Sideroblasts, RAEB=RA with
  Excess Blasts, RAEBT=RAEB in Transmission), Human Leukocyte (HLA)
  matching between donor and recipient, and donor/recipient gender.
  Negative values indicate MMF has lower incidence of cGVHD.
\label{Fig:forestplot}}
\end{figure}

While there appears to be little difference in the CIF of cGVHD
between the different GVHD prophylaxis strategies overall, it is also
worth examining whether this is consistent across subgroups.  We can
use the partial dependence functions to examine the difference in CIF
of cGVHD by 2~years between MTX and MMF in varous subgroups.  These
are shown as a forest plot in Figure~\ref{Fig:forestplot}.  These are
generally consistent with the overall findings, with most subgroups
showing posterior mean differences of less than~5\% in the 2~year CIF of
cGVHD, and a few showing differences of up to~7\%.

The {\bf BART} package can also be used to provide
predictions of the difference in cumulative incidence between the GVHD
prophylaxis regimens for each individual.  These are shown in
Figure~\ref{Fig:forestplot}(b), and show substantially more variability in
the individual predictions compared to the subgroup mean predictions, as expected.

%{\bf STILL TO DO: PLOT OF QUANTILE OF CIF VS. AGE}
%{\bf SHOULD WE DO SOMETHING LIKE THIS}
Finally, we examined the variable selection probabilities from fitting
the DART model to this data set, to identify which variables have the
highest posterior probabilities of being selected in the trees.  Only
five variables had at least a 5\% mean posterior probability of being
selected; these were, in order, time (48\%), use of MMF as GVHD
prophlaxis (7\%), use of in vivo T-cell depletion (6\%), use of
Flu/Mel conditioning (6\%), and AML patients in Primary Induction
Failure or Relapse (6\%).  The first four of these were all selected
in at least one of the trees in at least 90\% of the posterior
samples, while the last one was selected in 74\% of the posterior
samples.  None of the other variables were selected as consistently in
at least one of the trees.

\section{Conclusion}

In this article, we have proposed a novel approach for flexible
modeling of competing risks data using BART.  The model handles a
number of complexities in modeling, including nonlinear functions of
covariates, interactions, high-dimensional paramater spaces, and
nonproportional hazards (cause-specific or subdistribution).  It has
excellent prediction performance as a nonparametric ensemble
prediction model.

Our approach can be extended to handle missing data which is often
encountered in clinical research studies.  One approach, implemented
in {\bf bartMachine} \cite{KapeBlei14}, incorporates missing data
indicators into the training data set allowing for splits on the
missing indicators; this can improve performance under a pattern
mixture model framework.  An alternative approach uses sequential BART
models to impute the missing covariates \citep{XuDani16,DaniSing18}.

The methods proposed in this article can be computationally demanding,
due to the need to expand the data at a grid of event times; although,
Method~1 is less demanding of the two.  Nevertheless, we have found
that the computation times are competitive with Random Survival
Forests when you account for bootstrapping by RSF to obtain
uncertainty estimates.  Also, for large $P$, BART experiences only
modest increases in computation time, while RSF suffers from
substantial increases.  Our approach can be parallelized, since the
chains do not share information besides the data itself;
simultaneously performing calculations on $m$ chains can lead to
substantial improvements in processing time (nearly linear for small
$m$, but due to the burn-in penalty for each chain, diminishing
returns as $m$ increases further; see Amdahl's law of parallel computing
\cite{Amda67}).  The computational burden, particularly for large data
sets, can be reduced by coarsening the time scale so that the number
of grid points does not grow with $N$.  We are currently investigating
alternative models which do not require expansion of the data at a
grid of event times.

Our formulation allows for the use of ``off-the-shelf'' BART software
based on binary outcomes after restructuring the data as described.
Furthermore, we have incorporated the competing risks BART models into
our state-of-the-art {\bf BART} R package \cite{McCuSpar18} which is
publicly available on the Comprehensive R Archive Network (CRAN),
\url{https://cran.r-project.org}, and distributed under the GNU
General Public License.

\section{Acknowledgements}

Funding for this research was provided in part
by the Advancing Healthier Wisconsin Research and Education Program
%This work is supported in part by the Advancing a Healthier Wisconsin
%Endowment 
at the Medical College of Wisconsin.

\bibliographystyle{plain}
\bibliography{bibref} 

\end{document}